\documentclass[10pt,journal,twoside]{IEEEtran}
\usepackage{multirow}
\usepackage{amssymb}
\usepackage[dvips]{graphicx}
\usepackage{amsmath}
\usepackage{amsthm}
\usepackage{latexsym,bm}
\usepackage{color}
\usepackage{subfigure}
\usepackage{longtable}
\usepackage{accents}
\usepackage{cite}
\usepackage{enumerate}
\usepackage{arydshln}
\usepackage{slashbox}
\usepackage{float}
\usepackage[linesnumbered,ruled,vlined]{algorithm2e}
\usepackage{booktabs}
\usepackage{stfloats}
\usepackage{graphicx}
\usepackage{bm}
\usepackage{mathrsfs}

   \def\bu{{\mathbf{u}}}
   
\def\bc{{\mathbf{c}}}   \def\bw{{\mathbf{w}}}
  \def\br{{\mathbf{r}}} \def\bx{{\mathbf{x}}}
   \def\by{{\mathbf{y}}}
   \def\bz{{\mathbf{z}}}
 
 \def\bH{{\mathbf{H}}}  
\def\bB{{\mathbf{B}}}

  \def\bT{{\mathbf{T}}} \def\bZ{{\mathbf{Z}}}

\makeatletter
\def\widebar{\accentset{{\cc@style\underline{\mskip10mu}}}}
\def\Widebar{\accentset{{\cc@style\underline{\mskip8mu}}}}
\makeatother

\theoremstyle{plain}

\theoremstyle{definition}
\theoremstyle{definition} 

\begin{document}

\title{\huge Asymmetric Dual-Mode Constellation and Protograph LDPC Code Design for Generalized Spatial MPPM Systems\\
\thanks{
This work was supported in part by the NSF of China under Grant 62071131, the Guangdong Basic and Applied Basic Research Foundation under Grant 2022B1515020086, in part by the Open Research Fund of the State Key Laboratory of Integrated Services Networks under Grant ISN22-23, the International Collaborative Research Program of Guangdong Science and Technology Department under Grant 2022A0505050070, the Industrial Research and Development Project of Haoyang Electronic Company Ltd., under Grant 2022440002001494, and the National Research Foundation, Singapore University of Technology and Design under its Future Communications Research \& Development Programme ``Advanced Error Control Coding for 6G URLLC and mMTC'' Grant No. FCP-NTU-RG-2022-020. {\em(Corresponding author: Yi~Fang.)}}
\thanks{Liang Lv, Yi Fang, and Lin Dai are with the School of Information Engineering, Guangdong University of Technology, Guangzhou 510006, China, and also with the State Key Laboratory of Integrated Services Networks, Xidian University, Xi'an 710126, China (e-mail: lianglv0206@163.com; fangyi@gdut.edu.cn; bs0109dl@163.com). }
\thanks{Yonghui Li is with the School of Electrical and Information Engineering, The University of Sydney, Sydney, NSW 2006, Australia (e-mail: yonghui.li@sydney.edu.au).}
\thanks{Mohsen Guizani is with the Department of Machine Learning, Mohamed Bin Zayed University of Artificial Intelligence (MBZUAI), Abu Dhabi, UAE (e-mail: mguizani@ieee.org).}}

\author{
{Liang Lv, Yi Fang, \emph{Senior Member, IEEE}, Lin Dai, Yonghui Li, \emph{Fellow, IEEE}, and Mohsen Guizani, \emph{Fellow, IEEE}}\vspace{-7mm}}

\maketitle

\begin{abstract}
To achieve reliable and efficient transmissions in free-space optical (FSO) communication, this paper designs a new protograph low-density parity-check (PLDPC) coded generalized spatial multipulse position modulation (GSMPPM) system over weak turbulence channels. Specifically, we investigate the PLDPC code, generalized space shift keying (GSSK) modulation, and MPPM constellation. First, we propose a type of novel GSMPPM constellations that intelligently integrates the GSSK into MPPM, referred to as \emph{asymmetric dual-mode (ADM) constellations}, so as to improve the performance of the PLDPC-coded GSMPPM system. Furthermore, exploiting a protograph extrinsic information transfer (PEXIT) algorithm, we construct a type of improved PLDPC code, referred to as \emph{I-PLDPC code}, which outperforms the existing PLDPC codes over weak turbulence channels. Analytical and simulation results show that the proposed ADM constellations and the proposed I-PLDPC code can obtain noticeable performance gains over their counterparts. Therefore, the proposed PLDPC-coded GSMPPM system with ADM constellations is competent to satisfy the high-reliability requirement for FSO applications.
\end{abstract}


\vspace{-2mm}
\section{Introduction}\label{sect:Introduction}
Free-space optical (FSO) communication has been widely used for ground-to-satellite communication systems due to the high transmission rate and high reliability \cite{9138713,9319151,9970371}. In practical applications, due to the effect of air refractive index \cite{5439306,8438298}, atmospheric turbulence (also known as scintillation) is generated in FSO communications. The atmospheric turbulence causes random fluctuations of the amplitude and phase for the received signal at the receiver, which seriously deteriorate the system performance \cite{6844864,10026614}. To quantify the effect of atmospheric turbulence, several mathematical models have been proposed to describe the distribution of turbulence fading in FSO communication systems. For example, the lognormal distribution has been used to describe the weak turbulence channel in FSO communication systems \cite{9551206,8959173}.

The weak turbulence channel is one of the most typical channel models to describe FSO communication scenarios, which have attracted significant research interest in the past two decades \cite{1221769,9099546,8744610,8086218}.
In the FSO communication systems based on weak turbulence, the optical signal uses intensity modulation with direct detection (IM/DD)\cite{9321161}, which mainly adopts pulse position modulation (PPM) \cite{6581873} or on-off keying modulation (OOK) \cite{8862937}. Although OOK can provide high bandwidth efficiency, it suffers from low energy efficiency and high synchronization complexity
at the receiver \cite{5259916}. As an alternative, PPM can improve the energy efficiency by increasing the order of modulation \cite{6581873}.
However, the improvement of energy efficiency is obtained at the cost of bandwidth efficiency, which leads to a limited transmission capacity. As a variant of PPM, multipulse position modulation (MPPM) has been proposed to improve the bandwidth efficiency by utilizing multiple pulsed slots during each symbol transmission \cite{16882}. In the past decade, researchers have conceived various designs of MPPM constellations in FSO communication systems.
In \cite{5439306}, a Gray labeling search (GLS) algorithm which considers the Hamming distance between adjacent symbols has been proposed to construct a new type of MPPM constellations. Furthermore, an MPPM constellation subset selection (MCSS) algorithm has been proposed \cite{7833038}, in which the MPPM symbols are continuously added to the constellation subset by maximizing the Hamming distance between two adjacent symbols. Through this way, an MPPM constellation with the size of a power of two can be generated.
The above two MPPM constellations have been only employed in single-input-single-output (SISO) technique, which cannot attain spatial diversity.

To tackle the above issue, multiple-input multiple-output (MIMO) technique has been introduced in FSO communication systems \cite{9790303}. Space shift keying (SSK) \cite{5165332} is a spatial modulation technique that utilizes transmit antennas with distinct path characteristics to distinguish the transmitted symbols. To be specific, in\cite{6241395}, a low-complexity spatial pulse position modulation (SPPM) scheme that combines SSK with PPM has been proposed. In \cite{7881027}, the authors propose a generalized spatial pulse position modulation (GSPPM) scheme by utilizing the pulse inversion technique. The GSPPM scheme requires a DC bias to achieve pulse inversion, which adds additional power consumption. Moreover, the GSPPM scheme employs a single pulse position modulation (PPM), which cannot be further optimized in the signal-domain constellation. In \cite{sym11101232}, the authors have considered a modulation scheme based on a generalized spatial modulation (GSM) with multiple pulse amplitude and position modulation (MPAPM) in the visible light communication (VLC) systems. This work only exploits the conventional mapping rule of GSM and MPAPM, but does not design new bit-to-symbol mapping rules. Furthermore, in the GSM-MPAPM scheme, the number of activated LED groups must be a power of two, and thus all possible activated LED groups cannot be used sufficiently.
In addition, a spatial MPPM (SMPPM) scheme adopting SSK and MPPM has been considered in \cite{8647138}. So far, an in-depth investigation on GSMPPM is still lacking. Actually, a conventional GSMPPM constellation includes a spatial-domain constellation (i.e., effective activated antenna groups) and a signal-domain constellation (i.e., MPPM constellations). The sizes of both two types of constellations should be a power of two, and thus part of antenna groups and MPPM symbols keep idle. Therefore, how to design an efficient GSMPPM constellation using more antenna groups and MPPM symbols is a challenging problem.

An alternative method to improve the system performance is the employment of error-correction codes (ECCs). For example, Reed-solomon (RS) codes \cite{5259916,9503410} and Low-density parity-check (LDPC) codes \cite{9272655,photonics9050349} have been used in the FSO systems. Among all ECCs, a class of structured LDPC codes, i.e., protograph LDPC (PLDPC) codes \cite{thorpe2003low}, have received tremendous attention due to low complexity and close-to-capacity performance \cite{905935}. As well, the researchers have proposed a protograph extrinsic information transfer (PEXIT)\cite{7112076} algorithm for predicting the decoding thresholds of PLDPC codes in specific communication scenarios. With the aid of the PEXIT algorithm, the authors have constructed a PLDPC code (i.e., code-B) for Poisson channels \cite{6663748}. Nevertheless, the code-B PLDPC code may not perform well over weak turbulence channels due to the different distribution characteristics. Hence, it is crucial to design a type of PLDPC code tailored for the PLDPC-coded GSMPPM system over such a scenario.

Inspired by the above motivation, we make a comprehensive investigation on the PLDPC-coded GSMPPM systems over weak turbulence channels. Thus, the contributions of this work are summarized as follows.
\begin{enumerate}[1)]
  \item We propose a new GSMPPM scheme based on an asymmetric dual-mode (ADM) constellation search algorithm by removing the constraint condition that the sizes of spatial-domain constellation and signal-domain constellation must be a power of two.
  \item We analyze the constellation-constrained capacity of the proposed GSMPPM constellations in the case of using different MPPM slots.
  \item With the aid of the PEXIT algorithm, we construct an improved PLDPC code, referred to as {\em I-PLDPC code}, to achieve excellent decoding thresholds in the GSMPPM system.
  \item Analytical and simulation results reveal that the proposed PLDPC-coded ADM-aided GSMPPM system remarkably outperforms the existing counterparts over weak turbulence channels.
\end{enumerate}

The remainder of this paper is organized as follows. In Section~\ref{sect:Systems Model}, we propose the PLDPC-coded GSMPPM system and estimate its corresponding constellation-constrained capacity. In Section~\ref{sec:Design}, we present the design method of the ADM constellations. In Section~\ref{sec:Design_PLDPC}, we construct a new PLDPC code for the PLDPC-coded GSMPPM system. Simulation results are presented in Section~\ref{sec:simulation}, and the conclusion is made in Section~\ref{sec:conclusions}.

\section{Systems Model}\label{sect:Systems Model}

\subsection{Conventional GSMPPM} \label{subsec:Conventional}
Assuming that there are $N_{\rm{t}}$ transmit antennas and $N_{\rm{r}}$ receive antennas in the conventional GSMPPM. At each transmission instant, ${N_{\rm{a}}}~(2 \le {N_{\rm{a}}} \le {N_{\rm{t}}/2})$ out of $N_{\rm{t}}$ transmit antennas are chosen as an activated antenna group. Apparently, the number of all possible activated antenna groups is $N_{\rm s} = \binom{N_{\rm t}}{N_{\rm a}}$, where $\binom{N_{\rm t}}{N_{\rm a}} = N_{\rm t}! / N_{\rm a}!(N_{\rm t} - N_{\rm a})!$, and ``$!$'' denotes the factorial. However, the number of $N_{\rm{s}}$ is typically not a power of two, and thus all possible activated antenna groups cannot be utilized substantially. To ensure that the number of the spatial-domain constellation (i.e., effective activated antenna groups) is a power of two, one usually selects $N_{\rm e}$ effective activation antenna groups to transmit an $M$-ary MPPM symbol, where $N_{\rm e} = {{2}^{\lfloor \log _{2} {{N_{\rm{s}}}} \rfloor }}$ (i.e., closest to $N_{\rm s}$), and $\lfloor \cdot \rfloor$ denotes the floor function. Generally, each MPPM symbol consists of $l$ slots, with ${l_{\rm{a}}}\left( 2\le {l_{\rm{a}}} \le l/2 \right)$ pulsed slots and $\left( l-{l_{\rm{a}}} \right)$ non-pulsed slots. Thereby, the MPPM symbol can be characterized by a length-$l$ vector ${\rm{\bz}}_{p}=\left[ z_{p}^{1},z_{p}^{2},\ldots ,z_{p}^{l} \right]$, where $z_{p}^{q}\in \left\{ 0,1 \right\}$, $q=1,2,\ldots ,l$, $p=1,2,\ldots ,M$, ``$1$'' represents a pulsed slot, and ``$0$'' means a non-pulsed slot.

Accordingly, there are two kinds of constellations (i.e., spatial-domain constellation and signal-domain constellation) in the conventional GSMPPM. The size of the spatial-domain constellation depends on the number of effective activated antenna groups, while the size of the signal-domain constellation depends on the order of MPPM. Thus, every $m = \lfloor \log _{2} {{N_{\rm{s}}}} \rfloor + \lfloor \log _{2} {M_{\max}} \rfloor$ coded bits are divided into two parts at each transmission instant, the first ${m_{\rm{t}}}=\lfloor \log _{2} {{N_{\rm{s}}}} \rfloor$ coded bits are used to select the effective activated antenna group, while the remaining ${m_{\rm{s}}}= \lfloor \log _{2} {M_{\max}} \rfloor$ coded bits are mapped into an $M$-ary MPPM symbol.

\begin{figure}[t]
\centering
\hspace{0mm}
\includegraphics[width=3.5in,height=1.8in]{{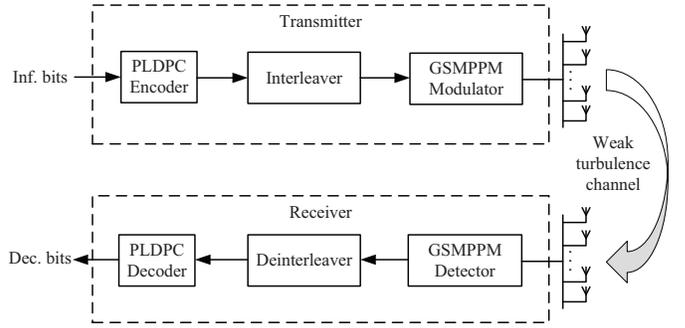}}
\caption{Block diagram of a PLDPC-coded GSMPPM system over a weak turbulence channel.}\vspace{-2mm}
\label{fig:system_model}
\end{figure}

\subsection{PLDPC-Coded GSMPPM System} \label{subsec:Systems}
The block diagram of a PLDPC-coded GSMPPM system is shown in Fig.~\ref{fig:system_model}, which has $N_{\rm t}$ transmit antennas and $N_{\rm r}$ receive antennas. Specifically, at the transmitter, a length-$k$ information-bit sequence ${\rm{\bu}}=\left\{{u_{1}},{u_{2}},\ldots ,{u_{k}} \right\}$ is first encoded by a PLDPC encoder to generate a length-$s$ codeword ${\rm{\bc}}=\left\{{{c}_{1}},{{c}_{2}},\ldots ,{{c}_{s}} \right\}$. Subsequently, $\rm{\bc}$ is passed to a GSMPPM modulator after permuted by a random interleaver. After that, every $m$ coded bits are mapped into GSMPPM symbol. Thus, a length-$(s/m)$ GSMPPM symbol sequence ${\rm{\bZ}}=\left\{ {{\rm{\bz}}_{1}},{{\rm{\bz}}_{2}},\ldots ,{{\rm{\bz}}_{s/m}} \right\}$ can be yielded. Note that we use $(N_{\rm t},N_{\rm r},N_{\rm a},l,l_{\rm a},M_{\rm s})$ to represent a GSMPPM constellation with size of $M_{\rm s} = 2^{m}$ in this paper. Finally, the GSMPPM symbol sequence ${\rm{\bZ}}$ is passed through a weak turbulence channel and each GSMPPM symbol ${\rm{\bz}}_{p}$ is converted into a transmission vector ${\rm{\bx}}$. The channel output $\rm{\by}$ can be written as
\begin{equation}\label{eq1}
    {\rm{\by}}=\frac{P_{\rm{t}}}{\sqrt{{N_{\rm{a}}}}} {\rm{\bH}} {\rm{\bx}} + {\rm{\bw}},
\end{equation}
where ${\rm{\by}} = [{\by}_1,{\by}_2,\ldots, {\by}_{N_{\rm r}} ]^{{\rm \bT}}$ denotes the received signal vector with size of $N_{\rm r} \times 1$ by all receive antennas, ${\by}_i = [{y}_i^{1}, {y}_i^{2}, \ldots, {y}_i^{l}] $ is the received signal in the $i$th receive antenna; ${\rm{\bx}} = [{\bx}_1,{\bx}_2,\ldots, {\bx}_{N_{\rm t}} ]^{{\rm \bT}}$ denotes the GSMPPM signal vector with size of $N_{\rm t} \times 1$ sent by transmit antennas, ${\bx}_j = [{x}_j^{1}, {x}_j^{2}, \ldots, {x}_j^{l}]$ is the GSMPPM signal sent by the $j$th transmit antenna; ${\rm{\bw}}$ denotes the noise matrix with size of ${N}_{\rm{r}} \times l$, in which each element is the real additive Gaussian noise with zero-mean and variance ${{\sigma}^{2}}={{{N}_{0}}}/{2}$, and ${N}_{0}$ is the noise power spectral density. Moreover, ${{P}_{\rm{t}}}= {P}_{\rm{a}}\cdot \gamma$ denotes the peak transmit power of the MPPM, where $P_{\rm{a}}$ is the average transmit power (all modulation patterns use the same ${P}_{\rm{a}}$), $\gamma = {1}/{\tau}$ is  the peak-to-average power ratio (PAPR), and $\tau ={{l}_{\rm{a}}}/{l}$ is the duty cycle of each MPPM symbol. In addition, ${\rm{\bH}}=({h}_{ij})$ denotes the channel coefficient matrix of size $N_{\rm r} \times N_{\rm t}$, where ${h}_{ij}$ is the channel coefficient from the $j$th transmit antenna to the $i$th receive antenna in the PLDPC-coded GSMPPM system over a weak turbulence channel. The channel coefficient ${h}_{ij}$ of the weak turbulence channel follows a lognormal distribution, and its probability density function (PDF) is given by \cite{9551206, 8959173}
\begin{equation}\label{eq2}
f({{h}_{ij}})=\frac{1}{2{{h}_{ij}}\sqrt{2\pi}{{\sigma }_{\rm{x}}}}\exp \left( -\frac{{{\left( \ln \left( {{h}_{ij}} \right)-2\mu  \right)}^{2}}}{8\sigma _{\rm{x}}^{2}} \right),
\end{equation}
where $\sigma_{\rm{x}}^{2}=0.25\ln\left( 1+\sigma _{\rm{I}}^{2} \right)$ is the log-amplitude variance, ${{\sigma}_{\rm{I}}}\left( \sigma _{\rm{I}}^{2} < 1 \right)$ is the channel scintillation index, and $\mu = -\sigma_{\rm{x}}^{2}$ is the log-amplitude mean \cite{8370053}. To ensure that the average power is not impacted by channel fading, the channel coefficients are normalized as $\mathbb{E}\left[h_{ij}^{2} \right]=1$, where $\mathbb{E}\left[ \cdot  \right]$ stands for the expectation function. The signal-to-noise ratio (SNR) can be expressed as\cite{6241395,7881027} SNR $ = \left( {{l}_{\rm{a}}}P_{\rm{t}}^{2} \right)/\left( 2 R m{{\sigma }^{2}} \right)$, where $R$ is the code rate.

At the receiver, the received signal $\by$ is detected by the max-sum approximation of log-domain maximum a-posterior probability (Max-log-MAP) algorithm \cite{7341061,5259916} in the GSMPPM detector. Then, the extrinsic log-likelihood ratios (LLRs) output from the GSMPPM detector are sent to a deinterleaver. After that, these extrinsic LLRs will be fed to a PLDPC decoder to perform belief-propagation (BP) \cite{661110, 8693678} decoding.

\subsection{Constellation-Constrained Capacity}\label{subsec:AMI}
Given the channel state information (CSI), the maximum rate of a reliable transmission can be determined by the average mutual information (AMI) \cite{9367298}. Assume that the selected GSMPPM symbol from the GSMPPM constellation is equiprobable, the constellation-constrained AMI of a code modulation (CM) scheme over a weak turbulence channel can be calculated as\cite{8237200}
\begin{equation}\label{eq3}
{{C}_{\rm{CM}}} = m-{{\mathbb{E}}_{\rm{\bx}, \rm{\by}, \rm{\bH}}}\left[ {{\log}_{2}}\frac{\sum\nolimits_{\rm{\br}\in \Omega }{p\left( \rm{\by} \mid \rm{\br},\rm{\bH} \right)}}{p\left( \rm{\by} \mid \rm{\bx}, \rm{\bH} \right)} \right],
\end{equation}
where
\begin{equation}\label{eq4}
  \begin{aligned}
    p\left( {\rm{\by}} \mid {\rm{\bx}}, \rm{\bH} \right)&= \prod_{i=1}^{N_{\rm r}} \prod_{q=1}^{l} p\left(y_{i}^{q} \mid {\rm{\bx}}, \rm{\bH}\right) \\
      &=\frac{\exp \left[-\frac{\sum\nolimits_{i=1}^{N_{\rm r}}  \sum\nolimits_{q = 1}^{l} \left(y_{i}^{q}- \sum\nolimits_{j=1}^{N_{\rm t}} \frac{P_{\rm t}}{\sqrt{N_{\rm{a}}}} {h_{ij}} x_{j}^{q}\right)^{2}} {2 \sigma^{2}}\right]}{{2 \pi \sigma^{2}}^\frac{l N_{\rm r}}{2}}.
  \end{aligned}
\end{equation}
In Eq.~\eqref{eq3}, $\Omega $ denotes a GSMPPM constellation, $y_{i}^{q}$ denotes the received signal of the $q$th slot at the $i$th receive antenna, $x_{j}^{q}$ denotes the signal transmitted by the $q$th slot at the $j$th transmit antenna, and $p({\rm{\by}} \mid {\rm{\bx}} , {\rm{\bH}} )$ is the probability density function (PDF) of the received signal vector $\rm{\by}$ under the conditions of the channel coefficient matrix $\rm{\bH}$ and the GSMPPM symbol vector $\rm{\bx}$. Therefore, in Eq.~\eqref{eq4}, the PDF of the elements $h_{ij}$ in the channel coefficient matrix $\rm{\bH}$ follows Eq.~\eqref{eq2}. In addition, the constellation-constrained AMI of a bit-interleaved coded modulation (BICM) scheme can be calculated as \cite{8237200}
\begin{equation}\label{eq5}
{{C}_{\rm{BICM}}} = m-\sum\limits_{k=1}^{m}{{{\mathbb{E}}_{b,{\rm{\by}},{\rm{\bH}}}}\left[ {{\log }_{2}}\frac{\sum\nolimits_{{\rm{\bx}}\in \Omega }{p\left( {\rm{\by}} \mid {\rm{\bx}},{\rm{\bH}} \right)}}{\sum\nolimits_{{\rm{\bx}}\in \Omega _{k}^{b}}{p\left( {\rm{\by}} \mid {\rm{\bx}},{\rm{\bH}} \right)}} \right]},
\end{equation}
where $\Omega_{k}^{b}$ denotes the subset of GSMPPM constellation $\Omega$ with the $k$th bit being $b\in \left\{ 0,1 \right\}$.

\section{Design of Proposed ADM constellations for PLDPC-coded GSMPPM System}\label{sec:Design}
\subsection{Proposed ADM Constellations}\label{subsec:ADM}
In the conventional GSMPPM modulation scheme, the $m_{\rm t}$ coded bits are used to select effective activated antenna groups, while the $m_{\rm s}$ coded bits are modulated by using MPPM symbols for effective activated antenna groups. Note that each effective activated antenna group shares the same MPPM constellation set $\Psi = \{\Psi_1,\Psi_2,\ldots,\Psi_M\}$. In practice, the number of all possible transmit antenna groups is $N_{\rm s}$, and the number of $N_{\rm{s}}$ is typically not a power of two. However, to ensure that the number of the spatial-domain constellation is a power of two, one usually selects $N_{\rm e}$ activated antenna groups as the effective activated antenna groups (i.e., the spatial-domain constellation) in the conventional GSMPPM system, where $N_{\rm e} = {{2}^{\lfloor \log _{2} {{N_{\rm{s}}}} \rfloor }}$ (i.e., closest to $N_{\rm s}$). Therefore, the activated antenna groups cannot be exploited sufficiently, i.e., $N_{\rm r}(N_{\rm r} = N_{\rm s} - N_{\rm e})$ activated antenna groups are idle. For instance, when $N_{\rm t} = 4 $ and $N_{\rm a} = 2$, $(1, 2)$, $(3, 4)$, $(1, 4)$, and $(2, 3)$ are selected as effective activated antenna groups, while the remaining activated antenna groups $(1, 3)$ and $(2, 4)$ are idle, where $1$, $2$, $3$ and $4$ denote the corresponding transmit antenna indices. Meanwhile, the full MPPM symbol set $\Phi$ consists of MPPM symbols with size of $M_{\max }=\binom{l}{l_{\rm a}}$, but the size of the full MPPM symbol set $\Phi$ is typically not a power of two. For this reason, the full MPPM symbol set $\Phi$ cannot be directly used for bit-to-symbol mapping in the GSMPPM system. In order to accommodate the transmission requirements, the number of MPPM symbols in the signal-domain constellation must be a power of two. Thus, one should select $M$ MPPM symbols from $\Phi$ as the signal-domain constellation, where $M = 2^{\lfloor \log_{2} {M_{\max}} \rfloor}$ (i.e., closest to $M_{\max}$). In the existing works, e.g., GLS \cite{5439306} and MCSS \cite{7833038} constellations, only $M$ MPPM symbols are selected from set $\Phi$ as an eligible sub-constellation set $\Psi \subset \Phi$ to perform the bit-to-symbol mapping, and $M_{\max} - M$ MPPM symbols remain idle.

\vspace{0mm}
 \begin{algorithm}[t]
    \caption{ADM Constellation Parameter Selection}\label{alg:1}
     \textbf{Initialization}: Given parameters ${N_{\rm t}}$, ${N_{\rm a}}$, $l$ and ${l_{\rm a}}$, calculate $N_{\rm s}$, $M_{\max }$, $m$, $M_{\rm s}$, and $M\leftarrow {{2}^{\lfloor \log _{2} {M_{\rm max}} \rfloor}}$. Set $Th = 0$, $i$ = $0$, ${N_{\rm add}} \leftarrow N_{\rm s}- N_{\rm e} $.\\

     \While{$Th == 0$} {Calculate ${M_{\rm A}}\leftarrow \left\lceil {M_{\rm s}}/{N_{\rm s}}\right\rceil+i$ and ${M_{\rm B}}$;\\

    \If {$({M_{\rm A}} > {M_{\rm B}})$ \&\& $({M_{\rm A}} + {M_{\rm B}}) \leq M_{\max}$ } {{\rm $Th$} = 1;\\ \textbf{continue}}

    {${N_{\rm add}} \leftarrow {N_{\rm add}} - 1$;}\\

    \If {${N_{\rm add}} == 0$}{Reset ${N_{\rm add}}$ and $i \leftarrow i + 1$;}}

    {\textbf{Finalization}: Output parameters: ${N_{\rm add}}$, $M_{\rm max}$, $M$, $m$, ${M_{\rm A}}$ and ${M_{\rm B}}$.}
 \end{algorithm}

To overcome the above weakness, we propose a type of asymmetric dual-mode (ADM) constellations by considering full MPPM symbols and all possible activated antenna groups as much as possible. The proposed ADM constellation scheme can eliminate the constraint condition that the sizes of the spatial-domain constellation and signal-domain constellation must be a power of two. Specifically, we first select $N_{\rm e}$ activated antenna groups (i.e., effective activated antenna groups), thus the number of the remaining activated antenna groups is $N_{\rm r}$. Subsequently, we define two MPPM sub-constellation sets ${\Psi }_{\rm A}$ and ${\Psi }_{\rm B}$. The MPPM symbols in the sub-constellation set ${\Psi }_{\rm A}$ with size of $M_{\rm A}$ ($M_{\rm A} < M$) are selected from the full MPPM symbol set $\Phi$, and the MPPM symbols in the constellation set ${\Psi }_{\rm B}$ with size of $M_{\rm B}$ ($M_{\rm B} < M_{\rm A}$) are selected from the $M_{\max} - M_{\rm A}$ idle MPPM symbols. Then, each effective activated antenna group sends the MPPM symbols in the sub-constellation set ${\Psi }_{\rm A}$. Also, we add the additional activated antenna groups to send the MPPM symbols in the constellation $\Psi_{\rm B}$, where the number of the additional activated antenna groups is $N_{\rm add}$, and the additional activated antenna groups are selected from the $N_{\rm r}$ idle activated antenna groups. The size of the spatial-domain constellation is $N_{\rm e} + N_{\rm add}$, and the size of the signal-domain constellation is $M_{\rm A} + M_{\rm B}$. As seen, neither the size of the spatial-domain constellation nor the size of the signal-domain constellation is a power of two.
More importantly, we consider the mapping relationship between MPPM symbols and coded bits in the ADM constellation. Actually, we propose a novel method to relabel MPPM symbols in the ADM constellation based on the maximum Hamming distance criterion. The detailed design steps of the proposed ADM constellations in the PLDPC-coded GSMPPM system are as follows.

\begin{enumerate}[1)]
  \item \textbf{\emph{ADM Constellation Parameter Selection}}:
   In order to utilize all activated antenna groups as much as possible, how to determine the number ${N_{\rm add}}$ of additional activated antenna groups is one of the most critical issues. Given system parameters $N_{\rm{t}}$, $N_{\rm{a}}$, $l$, ${l}_{\rm{a}}$ and $M_{\rm s}$, all possible activated antenna groups can be calculated as $N_{\rm s}$ and the number of full-MPPM symbols is $M_{\rm max }$. First, we randomly select $N_{\rm e}$ as effective activated antenna groups, thus the number of the remaining activated antenna groups is $N_{\rm r}$. The additional activated antenna groups are selected from the remaining activated antenna groups. We define two MPPM sub-constellation sets ${{\Psi }_{\rm A}}$ and ${{\Psi }_{\rm B}}$. As such, each effective activated antenna group sends GSMPPM symbols by using constellation set ${{\Psi }_{\rm A}}$ with size of ${M_{\rm A}}$, while each additional activated antenna group sends GSMPPM symbols by using constellation set ${{\Psi }_{\rm B}}$ with size of ${M_{\rm B}}$. Finally, the relationship between the ${M_{\rm A}}$, ${M_{\rm B}}$, and ${N_{\rm add}}$ can be represented as
  \begin{equation}\label{eq6}
  \begin{aligned}
   & {M_{\rm B}} = \lceil {\left({M_{\rm s}}-{ N_{\rm e}}\cdot {M_{\rm A}}\right)}/{N_{\rm add}} \rceil, \\
   & {\rm subject~to}:~{M_{\rm A}} > {M_{\rm B}}, \\
                     &~~~~~~~~~~\quad\quad {M_{\rm A}} + {M_{\rm B}} \leq M_{\max},
  \end{aligned}
  \end{equation}
  where the initial value of ${M_{\rm A}}$ is set to $\lceil{M_{\rm s}}/{N_{\rm s}}\rceil$ and the initial value of ${N_{\rm add}}$ is set to $N_{\rm s}- N_{\rm e} $. To elaborate further, the ADM constellation parameter selection is summarized in {\em Algorithm \ref{alg:1}}.

\item \textbf{\emph{ADM Constellation Formulation}}:
(i) The parameters ${N_{\rm add}}$, ${M_{\rm A}}$ and ${{M}_{\rm B}}$ of an ADM constellation are determined by {\em Algorithm \ref{alg:1}}. Considering an ADM GSMPPM constellation scheme with $M_{\rm s}$ MPPM symbols, there exist $M_{\rm s}$ labels in the corresponding constellations mapper, each label $\xi_{\alpha} = [\xi_{\alpha}^{1},\xi_{\alpha}^{2},\ldots,\xi_{\alpha}^{m}]$ consists of $m$ labeling bits, where $\alpha = 1,2,\ldots,M_{\rm s}$, $\xi_{\alpha}^{m} \in \{0,1\}$, and $\xi_{\alpha}$ is the binary representation of the index value $(\alpha - 1)$. To conveniently determine the MPPM symbols in sub-constellation set ${\Psi }_{\rm A}$, we first divide the $N_{\rm e} M_{\rm A}$ labels into $N_{\rm e}$ label subsets. The index values corresponding to the labels in the $\lambda$th label subset $\boldsymbol{\xi}_{\lambda}$ within ${\Psi }_{\rm A}$ belong to the interval $[\lambda M, (M_{\rm A}-1)+\lambda M]$ (see Table~\ref{tab:ADM-constellations-5bits}), and the remaining labels correspond to the MPPM symbols in sub-constellation set ${\Psi}_{\rm B}$, where $\lambda = 0, 1, \ldots,N_{\rm e}-1$.
Here, we take the $\lambda$th label subset $\boldsymbol{\xi}_{\lambda}$ as an example and select the corresponding sub-constellation set ${\Psi}_{\rm A}$. Especially, ${M_{\rm A}}$ MPPM symbols must be selected from a full MPPM symbol set $\Phi$ of size $M_{\max}$ to constitute an MPPM sub-constellation set ${{\Psi }_{{\rm A}_{p}}}$, where $p = 1,2,\cdots, \binom{M_{\rm max}}{M_{\rm A}}$. The $i$th MPPM sub-constellation symbol (i.e., ${\Psi }_{{{\rm A}_p},i} = [{\psi }_{{{\rm A}_{p}},i}^{1},{\psi }_{{{\rm A}_p},i}^{2},\ldots,{\psi }_{{{\rm A}_p},i}^{l}]$) can be marked by the $i$th label (i.e., ${{\xi}}_{\lambda,i} = [ {{\xi}}_{\lambda,i}^{1},{\xi}_{\lambda,i}^{2},\ldots ,{\xi}_{\lambda,i}^{m}]$) in $\boldsymbol{\xi}_{\lambda}$, where $i = 1,2,\ldots,M_{\rm A}$.
Further, the Hamming distance of arbitrary two different MPPM symbols (e.g., ${\Psi }_{{{\rm A}_{p}},i}$ and ${\Psi }_{{{\rm A}_{p}},j}$) is defined as $D^{ij}$. One can calculate the Hamming distance $D_{{\rm A}_{p},i} = \frac{1}{M_{\rm A}-1}\sum_{j = 1,i \neq j}^{M_{\rm A}} {D^{ij}}$ between the $i$th MPPM symbol ${{\Psi }_{{\rm A}_{p},i}}$ and the remaining $M_{\rm A}-1$ MPPM symbols. Then, the average Hamming distance of the sub-constellation ${\Psi}_{{\rm A}_{p}}$ can be obtain by $D_{{\rm{a}},p} = {\frac{1}{M_{\rm A}} \sum_{i=1}^{M_{\rm A}}{D_{{{\rm A}_p},i}} }$. Assume that $D_{{\rm{a}},p}$ is the maximum average Hamming distance $D_{\rm a}$, we select the sub-constellation ${{\Psi }_{{\rm A}_{p}}}$ for the next operation.
  For the sub-constellation set ${{\Psi }_{{\rm A}_{p}}}$, we relabel each MPPM symbol based on the maximum Hamming distance criterion. To be specific, when $D^{ij}=2l_a$ (i.e., the largest Hamming distance), we maximize the Hamming distance $d$ of arbitrary two different labels corresponding to two different MPPM symbols ${\Psi }_{{{\rm A}_{p}},i}$ and ${\Psi }_{{{\rm A}_{p}},j}$. If the case of $d = m$ does not exist, we consider $d = m-1,m-2,\ldots,1$ and so on. Based on the above operations, we can obtain a sub-constellation set ${\Psi }_{\rm A}$.\vspace{2mm}

\begin{table*}[t]
\tiny
\centering\vspace{-2mm}
\caption{Mapping relationship among labels, MPPM symbols and activated antenna groups for the proposed ADM constellations, where ${N}_{\rm{t}}=4$, ${N}_{\rm{r}}=4$, ${N}_{\rm{a}}=2$, and $\rho = 5$ bpcu (i.e., $l = 5$, $l_{\rm a} = 2$ and $l = 6$, $l_{\rm a} = 2$).}
\begin{tabular}{|c|c|c|c|c|c|c|c|c|c|}

\hline
Effective activated &Index&Label &${\Psi }_{\rm A}$  &${\Psi }_{\rm A}$  &Effective activated & Index &Label                    &${\Psi }_{\rm A}$&${\Psi }_{\rm A}$\\
antenna group      &value&$\boldsymbol{\xi}$  &$(l=5)$  &$(l=6)$      &antenna group     &value&$\boldsymbol{\xi}$      &$(l=5)$          &$(l=6)$\\
\hline\hline
          &$0$ &$0\ 0\ 0\ 0\ 0$ &$1\ 0\ 1\ 0\ 0$ &$1\ 1\ 0\ 0\ 0\ 0$  &        &$8$ &$0\ 1\ 0\ 0\ 0$  &$1\ 0\ 1\ 0\ 0$  &$1\ 1\ 0\ 0\ 0\ 0$\\
          &$1$ &$0\ 0\ 0\ 0\ 1$ &$0\ 1\ 1\ 0\ 0$ &$0\ 0\ 1\ 0\ 0\ 1$  &        &$9$ &$0\ 1\ 0\ 0\ 1$  &$0\ 1\ 1\ 0\ 0$  &$0\ 0\ 1\ 0\ 0\ 1$\\
$(1, 2)$  &$2$ &$0\ 0\ 0\ 1\ 0$ &$1\ 0\ 0\ 1\ 0$ &$0\ 1\ 0\ 1\ 0\ 0$  &$(3, 4)$&$10$ &$0\ 1\ 0\ 1\ 0$  &$1\ 0\ 0\ 1\ 0$  &$0\ 1\ 0\ 1\ 0\ 0$\\
          &$3$ &$0\ 0\ 0\ 1\ 1$ &$0\ 0\ 1\ 1\ 0$ &$0\ 0\ 1\ 1\ 0\ 0$  &        &$11$ &$0\ 1\ 0\ 1\ 1$  &$0\ 0\ 1\ 1\ 0$  &$0\ 0\ 1\ 1\ 0\ 0$\\
          &$4$ &$0\ 0\ 1\ 0\ 0$ &$1\ 0\ 0\ 0\ 1$ &$1\ 0\ 0\ 0\ 1\ 0$  &        &$12$ &$0\ 1\ 1\ 0\ 0$  &$1\ 0\ 0\ 0\ 1$  &$1\ 0\ 0\ 0\ 1\ 0$\\
          &$5$ &$0\ 0\ 1\ 0\ 1$ &$0\ 1\ 0\ 0\ 1$ &$0\ 0\ 0\ 0\ 1\ 1$  &        &$13$ &$0\ 1\ 1\ 0\ 1$  &$0\ 1\ 0\ 0\ 1$  &$0\ 0\ 0\ 0\ 1\ 1$\\
\hline
          &$16$ &$1\ 0\ 0\ 0\ 0$     &$1\ 0\ 1\ 0\ 0$     &$1\ 1\ 0\ 0\ 0\ 0$  &        &$24$ &$1\ 1\ 0\ 0\ 0$ &$1\ 0\ 1\ 0\ 0$ &$1\ 1\ 0\ 0\ 0\ 0$  \\
          &$17$ &$1\ 0\ 0\ 0\ 1$     &$0\ 1\ 1\ 0\ 0$     &$0\ 0\ 1\ 0\ 0\ 1$  &        &$25$ &$1\ 1\ 0\ 0\ 1$ &$0\ 1\ 1\ 0\ 0$ &$0\ 0\ 1\ 0\ 0\ 1$  \\
$(1, 4)$  &$18$ &$1\ 0\ 0\ 1\ 0$     &$1\ 0\ 0\ 1\ 0$     &$0\ 1\ 0\ 1\ 0\ 0$  &$(2, 3)$&$26$ &$1\ 1\ 0\ 1\ 0$ &$1\ 0\ 0\ 1\ 0$ &$0\ 1\ 0\ 1\ 0\ 0$ \\
          &$19$ &$1\ 0\ 0\ 1\ 1$     &$0\ 0\ 1\ 1\ 0$     &$0\ 0\ 1\ 1\ 0\ 0$  &        &$27$ &$1\ 1\ 0\ 1\ 1$ &$0\ 0\ 1\ 1\ 0$ &$0\ 0\ 1\ 1\ 0\ 0$ \\
          &$20$ &$1\ 0\ 1\ 0\ 0$     &$1\ 0\ 0\ 0\ 1$     &$1\ 0\ 0\ 0\ 1\ 0$  &        &$28$ &$1\ 1\ 1\ 0\ 0$ &$1\ 0\ 0\ 0\ 1$ &$1\ 0\ 0\ 0\ 1\ 0$ \\
          &$21$ &$1\ 0\ 1\ 0\ 1$     &$0\ 1\ 0\ 0\ 1$     &$0\ 0\ 0\ 0\ 1\ 1$  &        &$29$ &$1\ 1\ 1\ 0\ 1$ &$0\ 1\ 0\ 0\ 1$ &$0\ 0\ 0\ 0\ 1\ 1$ \\	
\hline
Additional activated &Index&Label &${\Psi }_{\rm B}$  &${\Psi }_{\rm B}$  &Additional activated&Index&Label                     &${\Psi }_{\rm B}$&${\Psi }_{\rm B}$\\
antenna group       &value&$\boldsymbol{\zeta}$  &$(l=5)$  &$(l=6)$      &antenna group     &value&$\boldsymbol{\zeta}$      &$(l=5)$          &$(l=6)$\\
\hline\hline
                 &$6$  &$0\ 0\ 1\ 1\ 0$     &$1\ 1\ 0\ 0\ 0$    &$0\ 1\ 0\ 0\ 0\ 1$   &        &$7$ &$0\ 0\ 1\ 1\ 1$ &$1\ 1\ 0\ 0\ 0$ &$0\ 1\ 0\ 0\ 0\ 1$ \\
$(1, 3)$         &$14$ &$0\ 1\ 1\ 1\ 0$     &$0\ 1\ 0\ 1\ 0$    &$0\ 0\ 1\ 0\ 1\ 0$   &$(2, 4)$&$15$ &$0\ 1\ 1\ 1\ 1$ &$0\ 1\ 0\ 1\ 0$ &$0\ 0\ 1\ 0\ 1\ 0$ \\
                 &$22$ &$1\ 0\ 1\ 1\ 0$     &$0\ 0\ 1\ 0\ 1$    &$0\ 0\ 0\ 1\ 0\ 1$   &        &$23$ &$1\ 0\ 1\ 1\ 1$ &$0\ 0\ 1\ 0\ 1$ &$0\ 0\ 0\ 1\ 0\ 1$ \\
                 &$30$ &$1\ 1\ 1\ 1\ 0$     &$0\ 0\ 0\ 1\ 1$    &$0\ 0\ 0\ 1\ 1\ 0$   &        &$31$ &$1\ 1\ 1\ 1\ 1$ &$0\ 0\ 0\ 1\ 1$ &$0\ 0\ 0\ 1\ 1\ 0$ \\
\hline
\end{tabular}\label{tab:ADM-constellations-5bits}
\end{table*}

(ii) In order to decrease the interference between the GSMPPM symbols in two different activated antenna groups, the same MPPM symbol cannot exist in both ${\Psi }_{\rm A}$ and ${\Psi }_{\rm B}$. Thus, we should remove all MPPM symbols belonging to ${\Psi }_{\rm A}$ from $\Phi$ (i.e., $\widebar{\Phi}_{\rm A} = \Phi /{\Psi }_{\rm A}$). Given a spectral efficiency $\rho$, the additional activated antenna groups will cause the truncation of the effective activated antenna groups corresponding to signal-domain constellations. For example, when $N_{\rm t}=4$, $N_{\rm a}=2$, and $M_{\rm A}=6$, two MPPM symbols corresponding to the labels $[00110]$ and $[00111]$ cannot combine with the effective activated antenna groups to form two GSMPPM symbols, it can only form two GSMPPM symbols with the additional activated antenna groups $(1,3)$ and $(2,4)$, respectively. Therefore, we reclassify the remaining $M_{\rm r}$ labels into $N_{\rm add}$ label subsets in a sequential order, where $M_{\rm r} = 2^{m}(M - M_{\rm A})$. When $M-M_{\rm A}\le N_{\rm add}$, the index value corresponding to the $\beta$th label in the $\mu$th label subset $\boldsymbol{\zeta}_{\mu}$ is $(\beta-1)M + M_{\rm A}+\mu$, where $\mu = 0,1,\ldots,N_{\rm add}-1$, $\beta = 1,2,\ldots,M_{\rm B}$ (see Table~\ref{tab:ADM-constellations-5bits}). Otherwise (i.e., $M-M_{\rm A} > N_{\rm add}$), the index values corresponding to the labels are divided evenly into $N_{\rm add}$ subsets in a sequential order (see Table~\ref{tab:ADM-constellations-6bits}).
Especially, each label in a label subset $\boldsymbol{\zeta}_{\mu}$ corresponds to an MPPM symbol in sub-constellation ${\Psi }_{\rm B}$\footnote{The selected sub-constellation sets ${\Psi}_{\rm A}$ and ${\Psi}_{\rm B}$ are also applicable to other label subsets $\boldsymbol{\xi}_{\lambda}$ and $\boldsymbol{\zeta}_{\mu}$, respectively.}. Taking the label subset $\boldsymbol{\zeta}_{\mu}$ as an example, we select $M_{\rm B}$ MPPM symbols from set $\widebar{\Phi}_{\rm A}$ to form a sub-constellation ${\Psi_{{\rm B},\eta}}$, where $\eta= 1,2,\ldots, \binom{M_{\rm max} - M_{\rm A}}{M_{\rm B}}$. Likewise, we calculate the average Hamming distance of the sub-constellation ${\Psi}_{{\rm B}_{\eta}}$, which can be obtain by $D_{{\rm{b}},\eta} = {\frac{1}{M_{\rm B}(M_{\rm B}-1)} \sum_{i=1}^{M_{\rm B}} \sum_{j=1, i \ne j}^{M_{\rm B}} {D^{ij}}} $. Assume that $D_{{\rm b},\eta}$ is the maximum average Hamming distance $D_{\rm b}$, we select the sub-constellation ${{\Psi }_{{\rm B}_{\eta}}}$ for the next operation. Then, when $D^{ij}=2l_a$ (i.e., the largest Hamming distance), we maximize the Hamming distance $d$ of arbitrary two different labels corresponding to two MPPM symbols ${\Psi }_{{{\rm B}_{\eta}},i}$ and ${\Psi }_{{{\rm B}_{\eta}},j}$. Finally, the subconstellation set ${\Psi }_{\rm B}$ can be obtained by the above operation.
\end{enumerate}

Based on the above two steps, a type of ADM constellations can be formulated. To illustrate further, the design method for ADM constellations is summarized in \emph{Algorithm \ref{alg:2}}.
For example, with the aid of \emph{Algorithm \ref{alg:2}}, the ADM constellations with spectral efficiencies $\rho = 5$ bits per channel use (bpcu) and $\rho = 6$ bpcu are shown in Tables~\ref{tab:ADM-constellations-5bits} and \ref{tab:ADM-constellations-6bits}, respectively.

{\em Remark:} The proposed ADM constellation can be directly applicable to other turbulence channels (e.g., the Malaga turbulence channel and the gamma-gamma turbulence channel), since the design criterion is independent of the fading distribution.

\begin{table*}[t]
\tiny
\centering\vspace{-2mm}
\caption{Mapping relationship among labels, MPPM symbols and activated antenna groups for the proposed ADM constellations, where ${N}_{\rm{t}}=4$, ${N}_{\rm{r}}=4$, ${N}_{\rm{a}}=2$, and $\rho = 6$ bpcu (i.e., $l = 7$, $l_{\rm a} = 2$ and $l = 8$, $l_{\rm a} = 2$).}
\begin{tabular}{|c|c|c|c|c|c|c|c|c|c|}

\hline
Effective activated &Index&Label &${\Psi }_{\rm A}$  &${\Psi }_{\rm A}$  &Effective activated&Index&Label                     &${\Psi }_{\rm A}$&${\Psi }_{\rm A}$\\
antenna group      &value&$\boldsymbol{\xi}$  &$(l=7)$  &$(l=8)$      &antenna group     &value&$\boldsymbol{\xi}$      &$(l=7)$          &$(l=8)$\\
\hline\hline
                 &$0$&$0\ 0\ 0\ 0\ 0\ 0$ &$0\ 0\ 0\ 0\ 0\ 1\ 1$ &$0\ 0\ 1\ 0\ 0\ 1\ 0\ 0$ &         &$16$&$0\ 1\ 0\ 0\ 0\ 0$ &$0\ 0\ 0\ 0\ 0\ 1\ 1$  &$0\ 0\ 1\ 0\ 0\ 1\ 0\ 0$\\
                 &$1$&$0\ 0\ 0\ 0\ 0\ 1$ &$0\ 0\ 1\ 0\ 0\ 0\ 1$ &$0\ 0\ 1\ 1\ 0\ 0\ 0\ 0$ &         &$17$&$0\ 1\ 0\ 0\ 0\ 1$ &$0\ 0\ 1\ 0\ 0\ 0\ 1$  &$0\ 0\ 1\ 1\ 0\ 0\ 0\ 0$\\
                 &$2$&$0\ 0\ 0\ 0\ 1\ 0$ &$0\ 0\ 0\ 1\ 0\ 1\ 0$ &$0\ 0\ 0\ 0\ 1\ 0\ 1\ 0$ &         &$18$&$0\ 1\ 0\ 0\ 1\ 0$ &$0\ 0\ 0\ 1\ 0\ 1\ 0$  &$0\ 0\ 0\ 0\ 1\ 0\ 1\ 0$\\
                 &$3$&$0\ 0\ 0\ 0\ 1\ 1$ &$1\ 0\ 0\ 1\ 0\ 0\ 0$ &$0\ 0\ 0\ 1\ 0\ 0\ 1\ 0$ &         &$19$&$0\ 1\ 0\ 0\ 1\ 1$ &$1\ 0\ 0\ 1\ 0\ 0\ 0$  &$0\ 0\ 0\ 1\ 0\ 0\ 1\ 0$\\
                 &$4$&$0\ 0\ 0\ 1\ 0\ 0$ &$0\ 1\ 0\ 0\ 0\ 1\ 0$ &$1\ 0\ 0\ 0\ 0\ 1\ 0\ 0$ &         &$20$&$0\ 1\ 0\ 1\ 0\ 0$ &$0\ 1\ 0\ 0\ 0\ 1\ 0$  &$1\ 0\ 0\ 0\ 0\ 1\ 0\ 0$\\
$(1, 2)$         &$5$&$0\ 0\ 0\ 1\ 0\ 1$ &$0\ 1\ 1\ 0\ 0\ 0\ 0$ &$1\ 0\ 0\ 1\ 0\ 0\ 0\ 0$ &$(3, 4)$ &$21$&$0\ 1\ 0\ 1\ 0\ 1$ &$0\ 1\ 1\ 0\ 0\ 0\ 0$  &$1\ 0\ 0\ 1\ 0\ 0\ 0\ 0$ \\
                 &$6$&$0\ 0\ 0\ 1\ 1\ 0$ &$1\ 0\ 0\ 0\ 0\ 1\ 0$ &$0\ 0\ 0\ 0\ 1\ 1\ 0\ 0$ &         &$22$&$0\ 1\ 0\ 1\ 1\ 0$ &$1\ 0\ 0\ 0\ 0\ 1\ 0$  &$0\ 0\ 0\ 0\ 1\ 1\ 0\ 0$\\
                 &$7$&$0\ 0\ 0\ 1\ 1\ 1$ &$1\ 1\ 0\ 0\ 0\ 0\ 0$ &$1\ 0\ 0\ 0\ 1\ 0\ 0\ 0$ &         &$23$&$0\ 1\ 0\ 1\ 1\ 1$ &$1\ 1\ 0\ 0\ 0\ 0\ 0$  &$1\ 0\ 0\ 0\ 1\ 0\ 0\ 0$\\
                 &$8$&$0\ 0\ 1\ 0\ 0\ 0$ &$0\ 0\ 0\ 0\ 1\ 0\ 1$ &$0\ 1\ 0\ 0\ 0\ 0\ 0\ 1$ &         &$24$&$0\ 1\ 1\ 0\ 0\ 0$ &$0\ 0\ 0\ 0\ 1\ 0\ 1$  &$0\ 1\ 0\ 0\ 0\ 0\ 0\ 1$\\
                 &$9$&$0\ 0\ 1\ 0\ 0\ 1$ &$0\ 0\ 1\ 0\ 1\ 0\ 0$ &$0\ 0\ 1\ 0\ 0\ 0\ 0\ 1$ &         &$25$&$0\ 1\ 1\ 0\ 0\ 1$ &$0\ 0\ 1\ 0\ 1\ 0\ 0$  &$0\ 0\ 1\ 0\ 0\ 0\ 0\ 1$\\
                 &$10$&$0\ 0\ 1\ 0\ 1\ 0$ &$0\ 0\ 0\ 1\ 1\ 0\ 0$&$0\ 1\ 0\ 0\ 0\ 0\ 1\ 0$ &         &$26$&$0\ 1\ 1\ 0\ 1\ 0$ &$0\ 0\ 0\ 1\ 1\ 0\ 0$  &$0\ 1\ 0\ 0\ 0\ 0\ 1\ 0$ \\

\hline
                 &$32$&$1\ 0\ 0\ 0\ 0\ 0$ &$0\ 0\ 0\ 0\ 0\ 1\ 1$  &$0\ 0\ 1\ 0\ 0\ 1\ 0\ 0$  &         &$48$ &$1\ 1\ 0\ 0\ 0\ 0$    &$0\ 0\ 0\ 0\ 0\ 1\ 1$  &$0\ 0\ 1\ 0\ 0\ 1\ 0\ 0$\\
                 &$33$&$1\ 0\ 0\ 0\ 0\ 1$ &$0\ 0\ 1\ 0\ 0\ 0\ 1$  &$0\ 0\ 1\ 1\ 0\ 0\ 0\ 0$  &         &$49$&$1\ 1\ 0\ 0\ 0\ 1$    &$0\ 0\ 1\ 0\ 0\ 0\ 1$  &$0\ 0\ 1\ 1\ 0\ 0\ 0\ 0$\\
                 &$34$&$1\ 0\ 0\ 0\ 1\ 0$ &$0\ 0\ 0\ 1\ 0\ 1\ 0$  &$0\ 0\ 0\ 0\ 1\ 0\ 1\ 0$  &         &$50$&$1\ 1\ 0\ 0\ 1\ 0$    &$0\ 0\ 0\ 1\ 0\ 1\ 0$  &$0\ 0\ 0\ 0\ 1\ 0\ 1\ 0$\\
                 &$35$&$1\ 0\ 0\ 0\ 1\ 1$ &$1\ 0\ 0\ 1\ 0\ 0\ 0$  &$0\ 0\ 0\ 1\ 0\ 0\ 1\ 0$  &         &$51$&$1\ 1\ 0\ 0\ 1\ 1$    &$1\ 0\ 0\ 1\ 0\ 0\ 0$  &$0\ 0\ 0\ 1\ 0\ 0\ 1\ 0$\\
                 &$36$&$1\ 0\ 0\ 1\ 0\ 0$ &$0\ 1\ 0\ 0\ 0\ 1\ 0$  &$1\ 0\ 0\ 0\ 0\ 1\ 0\ 0$  &         &$52$&$1\ 1\ 0\ 1\ 0\ 0$    &$0\ 1\ 0\ 0\ 0\ 1\ 0$  &$1\ 0\ 0\ 0\ 0\ 1\ 0\ 0$\\
$(1, 4)$         &$37$&$1\ 0\ 0\ 1\ 0\ 1$ &$0\ 1\ 1\ 0\ 0\ 0\ 0$  &$1\ 0\ 0\ 1\ 0\ 0\ 0\ 0$  &$(2, 3)$ &$53$&$1\ 1\ 0\ 1\ 0\ 1$    &$0\ 1\ 1\ 0\ 0\ 0\ 0$  &$1\ 0\ 0\ 1\ 0\ 0\ 0\ 0$\\
                 &$38$&$1\ 0\ 0\ 1\ 1\ 0$ &$1\ 0\ 0\ 0\ 0\ 1\ 0$  &$0\ 0\ 0\ 0\ 1\ 1\ 0\ 0$  &         &$54$&$1\ 1\ 0\ 1\ 1\ 0$    &$1\ 0\ 0\ 0\ 0\ 1\ 0$  &$0\ 0\ 0\ 0\ 1\ 1\ 0\ 0$\\
                 &$39$&$1\ 0\ 0\ 1\ 1\ 1$ &$1\ 1\ 0\ 0\ 0\ 0\ 0$  &$1\ 0\ 0\ 0\ 1\ 0\ 0\ 0$  &         &$55$&$1\ 1\ 0\ 1\ 1\ 1$    &$1\ 1\ 0\ 0\ 0\ 0\ 0$  &$1\ 0\ 0\ 0\ 1\ 0\ 0\ 0$\\
                 &$40$&$1\ 0\ 1\ 0\ 0\ 0$ &$0\ 0\ 0\ 0\ 1\ 0\ 1$  &$0\ 1\ 0\ 0\ 0\ 0\ 0\ 1$  &         &$56$&$1\ 1\ 1\ 0\ 0\ 0$    &$0\ 0\ 0\ 0\ 1\ 0\ 1$  &$0\ 1\ 0\ 0\ 0\ 0\ 0\ 1$\\
                 &$41$&$1\ 0\ 1\ 0\ 0\ 1$ &$0\ 0\ 1\ 0\ 1\ 0\ 0$  &$0\ 0\ 1\ 0\ 0\ 0\ 0\ 1$  &         &$57$&$1\ 1\ 1\ 0\ 0\ 1$    &$0\ 0\ 1\ 0\ 1\ 0\ 0$  &$0\ 0\ 1\ 0\ 0\ 0\ 0\ 1$\\
                 &$42$&$1\ 0\ 1\ 0\ 1\ 0$ &$0\ 0\ 0\ 1\ 1\ 0\ 0$  &$0\ 1\ 0\ 0\ 0\ 0\ 1\ 0$  &         &$58$&$1\ 1\ 1\ 0\ 1\ 0$	&$0\ 0\ 0\ 1\ 1\ 0\ 0$  &$0\ 1\ 0\ 0\ 0\ 0\ 1\ 0$\\
\hline
Additional activated &Index&Label &${\Psi }_{\rm B}$  &${\Psi }_{\rm B}$  &Additional activated&Index&Label                     &${\Psi }_{\rm B}$&${\Psi }_{\rm B}$\\
antenna group       &value&$\boldsymbol{\zeta}$  &$(l=7)$  &$(l=8)$      &antenna group   &value&$\boldsymbol{\zeta}$      &$(l=7)$          &$(l=8)$\\
\hline\hline
                 &$11$&$0\ 0\ 1\ 0\ 1\ 1$ &$0\ 0\ 1\ 0\ 0\ 1\ 0$    &$1\ 0\ 1\ 0\ 0\ 0\ 0\ 0$      &        &$43$&$1\ 0\ 1\ 0\ 1\ 1$   &$0\ 0\ 1\ 0\ 0\ 1\ 0$   &$1\ 0\ 1\ 0\ 0\ 0\ 0\ 0$ \\
                 &$12$&$0\ 0\ 1\ 1\ 0\ 0$ &$0\ 0\ 0\ 1\ 0\ 0\ 1$    &$0\ 1\ 0\ 1\ 0\ 0\ 0\ 0$      &        &$44$&$1\ 0\ 1\ 1\ 0\ 0$   &$0\ 0\ 0\ 1\ 0\ 0\ 1$   &$0\ 1\ 0\ 1\ 0\ 0\ 0\ 0$ \\
                 &$13$&$0\ 0\ 1\ 1\ 0\ 1$ &$0\ 0\ 1\ 1\ 0\ 0\ 0$    &$0\ 1\ 0\ 0\ 1\ 0\ 0\ 0$      &        &$45$&$1\ 0\ 1\ 1\ 0\ 1$   &$0\ 0\ 1\ 1\ 0\ 0\ 0$    &$0\ 1\ 0\ 0\ 1\ 0\ 0\ 0$ \\
                 &$14$&$0\ 0\ 1\ 1\ 1\ 0$ &$1\ 0\ 0\ 0\ 0\ 0\ 1$    &$0\ 1\ 1\ 0\ 0\ 0\ 0\ 0$      &        &$46$&$1\ 0\ 1\ 1\ 1\ 0$   &$1\ 0\ 0\ 0\ 0\ 0\ 1$    &$0\ 1\ 1\ 0\ 0\ 0\ 0\ 0$ \\
$(1, 3)$         &$15$&$0\ 0\ 1\ 1\ 1\ 1$ &$1\ 0\ 1\ 0\ 0\ 0\ 0$    &$0\ 0\ 1\ 0\ 1\ 0\ 0\ 0$      &$(2, 4)$&$47$&$1\ 0\ 1\ 1\ 1\ 1$   &$1\ 0\ 1\ 0\ 0\ 0\ 0$    &$0\ 0\ 1\ 0\ 1\ 0\ 0\ 0$ \\
                 &$27$&$0\ 1\ 1\ 0\ 1\ 1$ &$0\ 0\ 0\ 0\ 1\ 1\ 0$    &$1\ 0\ 0\ 0\ 0\ 0\ 1\ 0$      &        &$59$&$1\ 1\ 1\ 0\ 1\ 1$   &$0\ 0\ 0\ 0\ 1\ 1\ 0$    &$1\ 0\ 0\ 0\ 0\ 0\ 1\ 0$ \\
                 &$28$&$0\ 1\ 1\ 1\ 0\ 0$ &$0\ 1\ 0\ 0\ 0\ 0\ 1$    &$0\ 0\ 0\ 1\ 0\ 1\ 0\ 0$      &        &$60$&$1\ 1\ 1\ 1\ 0\ 0$   &$0\ 1\ 0\ 0\ 0\ 0\ 1$    &$0\ 0\ 0\ 1\ 0\ 1\ 0\ 0$ \\
                 &$29$&$0\ 1\ 1\ 1\ 0\ 1$ &$0\ 1\ 0\ 1\ 0\ 0\ 0$    &$0\ 0\ 0\ 1\ 0\ 0\ 0\ 1$      &        &$61$&$1\ 1\ 1\ 1\ 0\ 1$   &$0\ 1\ 0\ 1\ 0\ 0\ 0$    &$0\ 0\ 0\ 1\ 0\ 0\ 0\ 1$ \\
                 &$30$&$0\ 1\ 1\ 1\ 1\ 0$ &$0\ 1\ 0\ 0\ 1\ 0\ 0$    &$0\ 0\ 0\ 0\ 0\ 1\ 0\ 1$      &        &$62$&$1\ 1\ 1\ 1\ 1\ 0$   &$0\ 1\ 0\ 0\ 1\ 0\ 0$    &$0\ 0\ 0\ 0\ 0\ 1\ 0\ 1$ \\
                 &$31$&$0\ 1\ 1\ 1\ 1\ 1$ &$1\ 0\ 0\ 0\ 1\ 0\ 0$    &$0\ 0\ 0\ 0\ 0\ 0\ 1\ 1$      &        &$63$&$1\ 1\ 1\ 1\ 1\ 1$   &$1\ 0\ 0\ 0\ 1\ 0\ 0$    &$0\ 0\ 0\ 0\ 0\ 0\ 1\ 1$ \\
\hline
\end{tabular}\label{tab:ADM-constellations-6bits}
\end{table*}

 \begin{algorithm}[t]
  \caption{ADM Constellation Formulation}\label{alg:2}
     \textbf{Initialization}: Given parameters ${N_{\rm add}}$, ${M_{\rm A}}$, ${M_{\rm B}}$, $M_{\rm max}$, $M$, $m$, $l$ and ${{l}_{\rm a}}$, generate a full MPPM symbol set $\Phi$ of size $M_{\max}$. Set $D_{{\rm a}} = 0$, $D_{{\rm b}} = 0$, Label = $0$.  \\

     \For{$p = 1, 2, \ldots, \binom{M_{\rm max}}{M_{\rm A}}$} {Generate a sub-constellation ${\Psi }_{{\rm A}_p}$ and calculate $D_{{\rm{a}},p}$;\\
          \If{$D_{{\rm{a}},p} > D_{{\rm a}}$}{Label $\leftarrow p$; $D_{{\rm a}} \leftarrow D_{{\rm{a}},p}$;}}

     \For{$i = 1, 2, \ldots, {M_{\rm A}}$}{\For{$j = i+1, i+2, \ldots, {M_{\rm A}}$ } { \If{$D^{ij}==2l_{\rm a}$}{Maximize the Hamming distance $d$ between two different labels in $\boldsymbol{\xi}_{\lambda}$; }} }

     ${\Psi }_{\rm A} \leftarrow {\Psi }_{{\rm A},{\rm Label}}$.\\
     Remove MPPM symbols in ${\Psi }_{\rm A}$ from $\Phi$, i.e., $\Phi$ = $\Phi /{\Psi }_{\rm A}$.\\
     \For{$\eta = 1, 2, \ldots, \binom{M_{\rm max} - M_{\rm A}}{M_{\rm B}}$}{Generate a sub-constellation ${\Psi }_{{\rm B}, \eta}$ and calculate $D_{{\rm b},\eta}$;\\
         \If{$D_{{\rm{b}},\eta} > D_{{\rm b}}$}{Label $\leftarrow \eta$; $D_{{\rm b}} \leftarrow D_{{\rm{b}},\eta}$;}}
     \For{$i = 1, 2, \ldots, {M_{\rm B}}$}{ \For{$j = i+1, i+2, \ldots, {M_{\rm B}}$ }{\If{$D^{ij}==2l_{\rm a}$}{Maximize the Hamming distance $d$ between two different labels in $\boldsymbol{\zeta}_{\mu}$;}}}
     ${\Psi }_{\rm B} \leftarrow {\Psi }_{{\rm B},{\rm Label}}$.\\
    {\textbf{Finalization}: Output ${\Psi }_{\rm A}$, ${\Psi }_{\rm B}$.}
 \end{algorithm}

\subsection{Performance Analysis}\label{subsec:Capacity}
\begin{figure}[t]
\centering
\subfigure[\hspace{-0.5cm}]{\label{fig:Capacity-5slots}
\includegraphics[width=3.2in,height=2.56in]{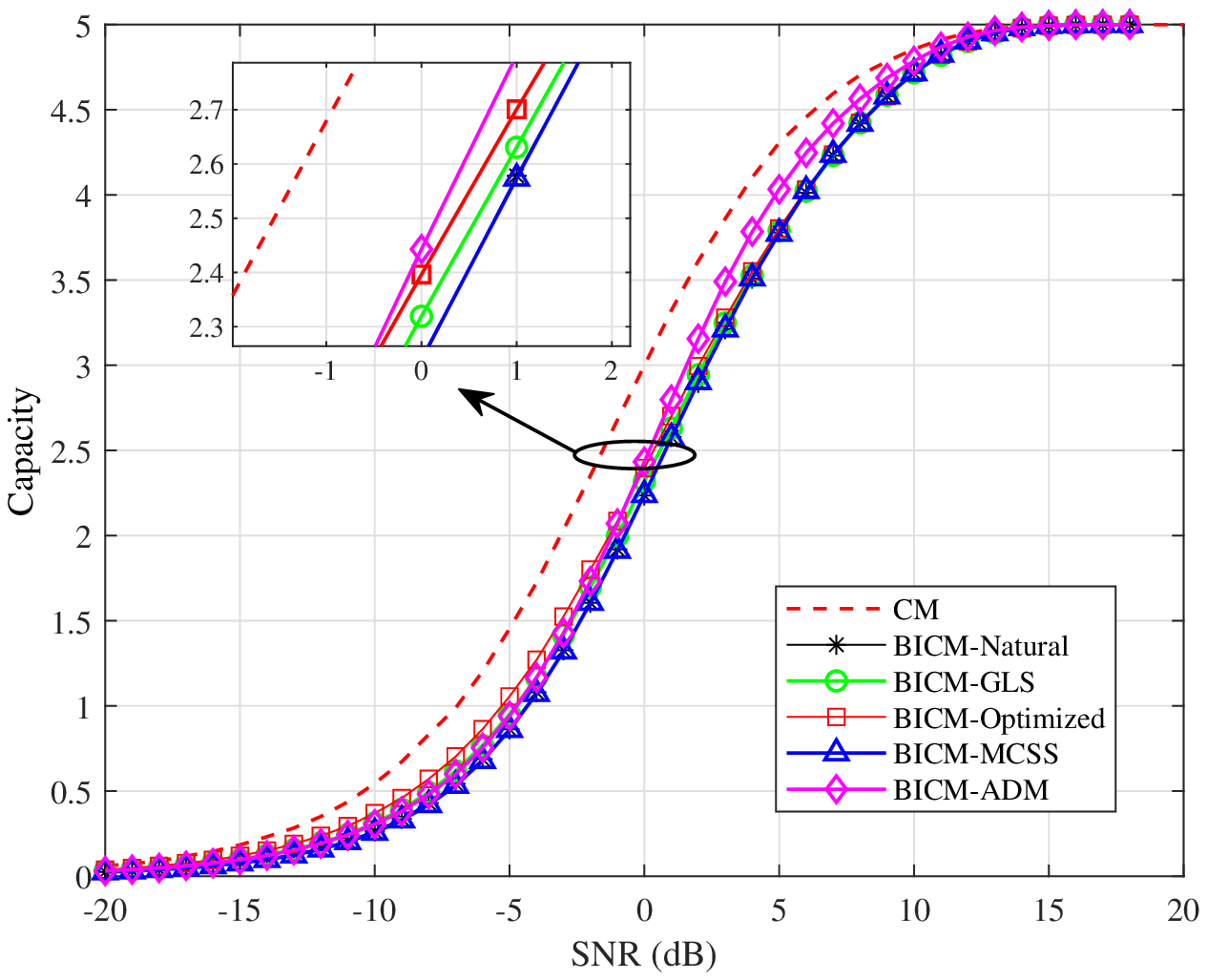}}
\subfigure[\hspace{-0.5cm}]{\label{fig:Capacity-6slots}
\includegraphics[width=3.2in,height=2.56in]{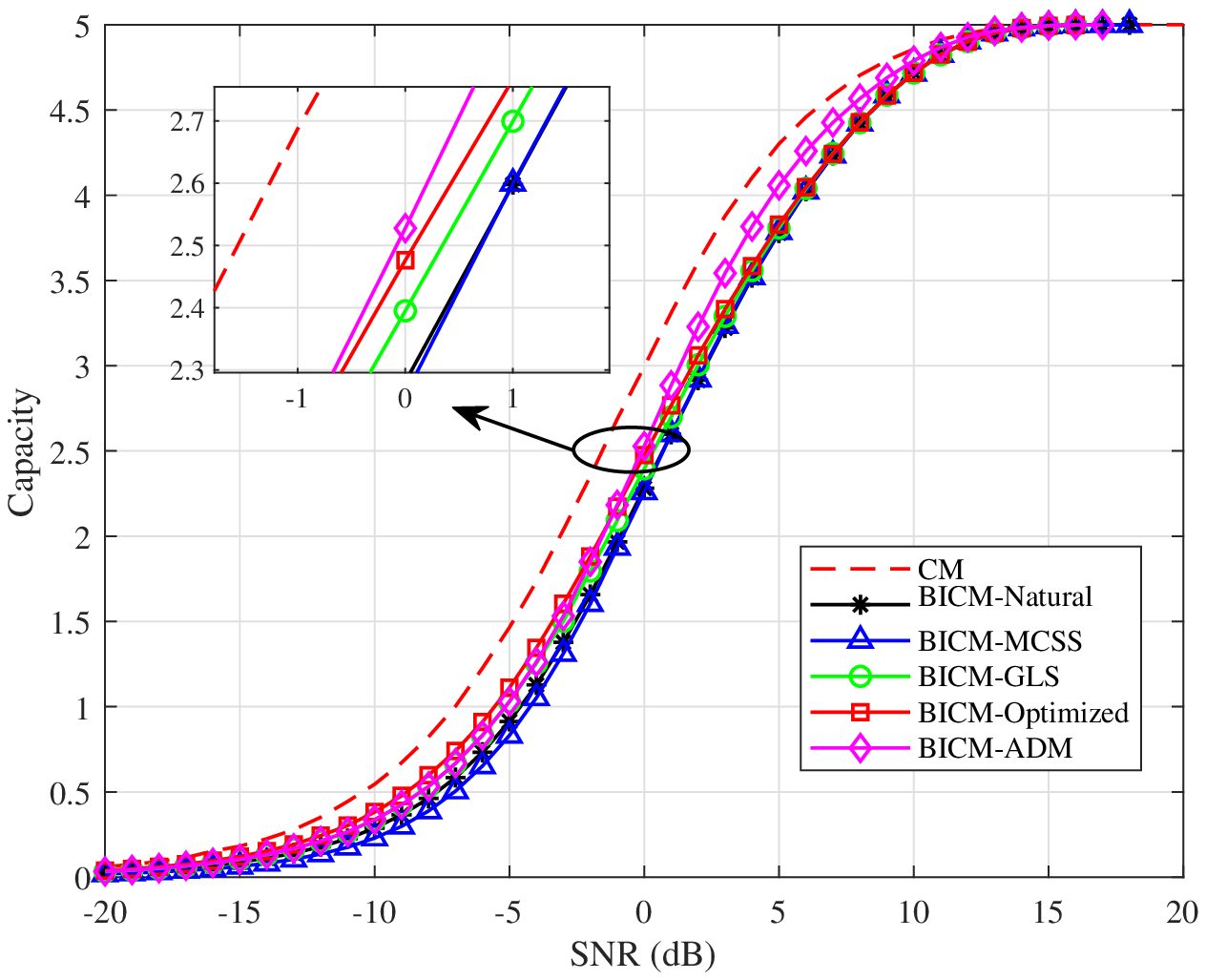}}
\vspace{-0.2cm}
\caption{Capacities of the PLDPC-coded GSMPPM systems with the proposed ADM constellation, the optimized constellation, natural constellation, MCSS constellation, and GLS constellation: (a) $(4,4,2,5,2,32)$-GSMPPM with $\rho=5$ bpcu; (b) $(4,4,2,6,2,32)$-GSMPPM with $\rho=5$ bpcu.}\vspace{-5mm}
\label{fig:Capacity-5bits}
\end{figure}

\begin{figure}[t]
\centering
\subfigure[\hspace{-0.5cm}]{\label{fig:Capacity-7slots}
\includegraphics[width=3.2in,height=2.56in]{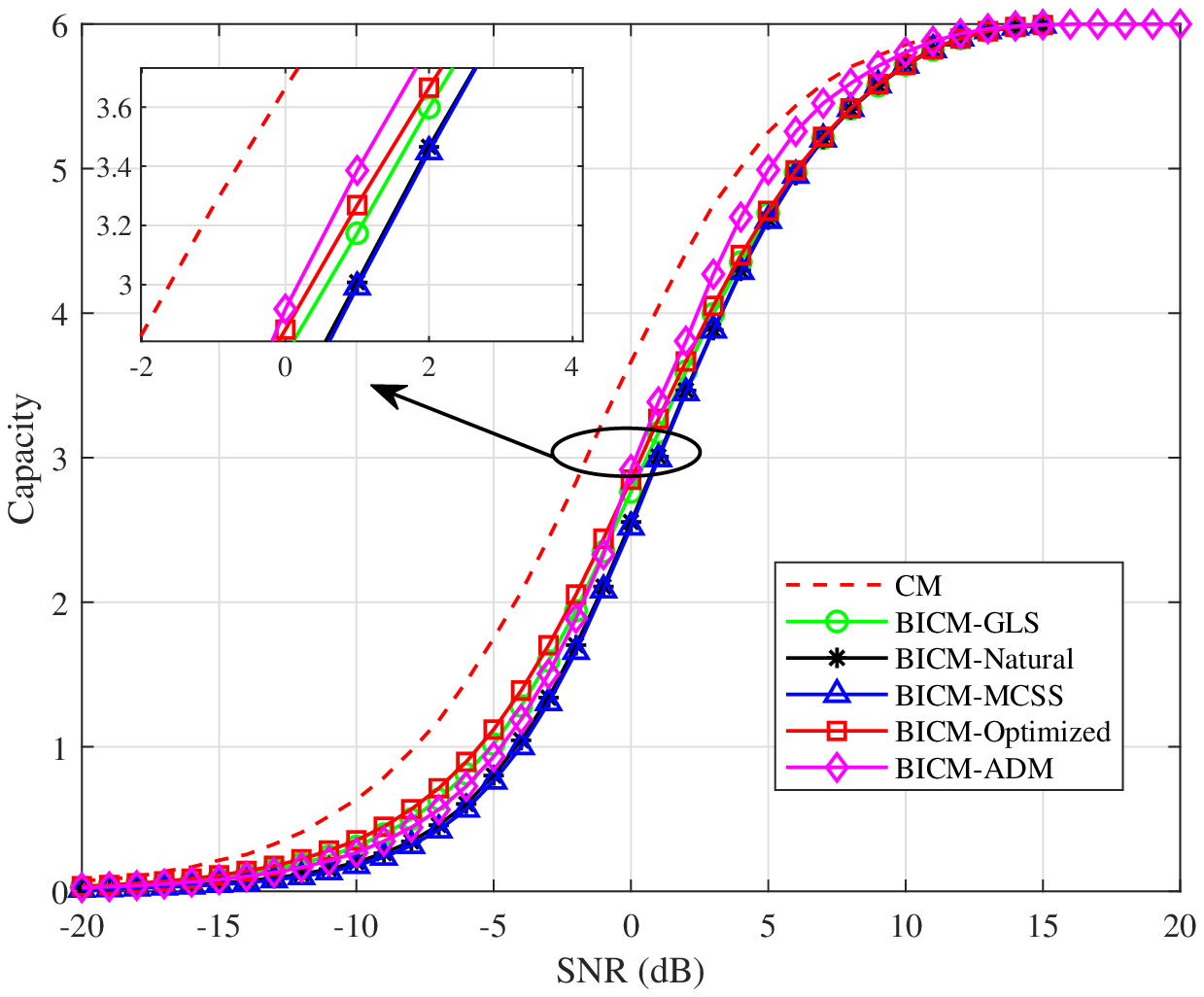}}
\subfigure[\hspace{-0.5cm}]{\label{fig:Capacity-8slots}
\includegraphics[width=3.2in,height=2.56in]{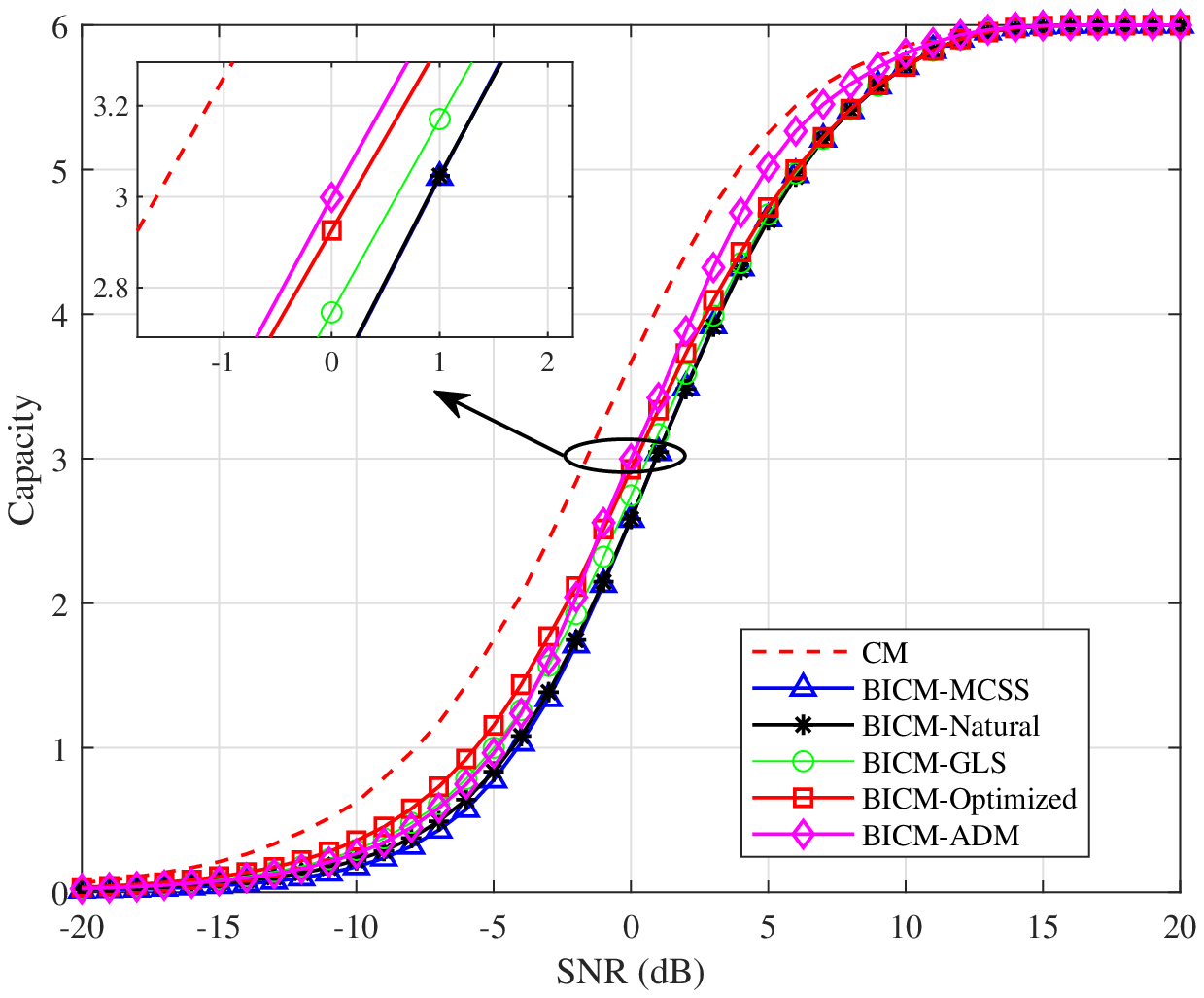}}
\vspace{-0.2cm}
\caption{Capacities of the PLDPC-coded GSMPPM systems with the proposed ADM constellation, the optimized constellation, natural constellation, MCSS constellation, and GLS constellation: (a) $(4,4,2,7,2,64)$-GSMPPM with $\rho=6$ bpcu; (b) $(4,4,2,8,2,64)$-GSMPPM with $\rho=6$ bpcu.}\vspace{-5mm}
\label{fig:Capacity-6bits}
\end{figure}

To verify the effectiveness of the proposed ADM constellations, the CM and BICM capacities of different GSMPPM constellations over the weak turbulence channel can be estimated by using the constellation-constrained capacity analysis method in Section~\ref{subsec:AMI}. In the PLDPC-coded GSMPPM systems, we calculate the constellation-constrained capacities of the proposed ADM constellations, the optimized constellations, natural constellation \cite{liu2009improved}, MCSS constellation \cite{7833038}, and GLS constellation \cite{5439306} with $\rho=5$ bpcu and $\rho=6$ bpcu are illustrated in Fig.~\ref{fig:Capacity-5bits} and Fig.~\ref{fig:Capacity-6bits}, respectively, where ${N_{\rm{t}}} = 4$, ${N_{\rm{r}}} = 4$, ${N_{\rm{a}}} = 2$ and ${{\sigma}_{\rm{x}}} = 0.3$.{\footnote{The optimized constellations are special cases of the proposed ADM constellations, which can be obtained by Algorithm 2 with the assumption of $N_{\rm add} = 0$, $M_{\rm A} = M$ and $M_{\rm B} = 0$. Based on the above parameter setting, the size of the MPPM symbol set for the optimized constellation is the same as those in the GLS and MCSS constellations. In this case, the optimized constellation can be used for performance comparison with the MPPM constellations in \cite{7833038,5439306}. The results illustrate that the optimized constellation is superior to the GLS and MCSS constellations, which further verifies the effectiveness of our proposed design.}}
Note that the quantization of turbulence in the weak turbulence channel is implemented by setting  the value of the log-amplitude variance $\sigma_{\rm x}$\cite{5259916}. For the case of $(4,4,2,5,2,32)$ GSMPPM scheme with $\rho=5$ bpcu, in Fig.~\ref{fig:Capacity-5slots}, one can see that the proposed ADM constellation and the optimized constellation can obtain much larger capacities than the other three constellations when the code rate $R > 0.35$ (i.e., $R={{{\mathcal{C}}_{\rm{BICM}}}}/{m}$).{\footnote{In order to facilitate the comparison with other GSMPPM mapping schemes, we refer to the mapping scheme formulated by transmit antenna groups and MPPM constellations as the $(N_{\rm t}, N_{\rm r}, N_{\rm a}, l, l_{\rm a}, M_{\rm s})$ GSMPPM scheme.}} In particular, the proposed ADM constellation is closest to the CM capacity. In Fig.~\ref{fig:Capacity-6slots}, we can observe similar results for $(4, 4, 2, 6, 2, 32)$ GSMPPM scheme. Likewise, the same phenomenon can be observed when the spectral efficiency $\rho=6$ bpcu (see Fig.~\ref{fig:Capacity-6bits}).
Therefore, it can be concluded that the proposed ADM constellations are able to obtain better performance with respect to the existing counterparts in the PLDPC-coded GSMPPM systems.

Moreover, we analyze the energy efficiency of the proposed GSMPPM scheme, the existing GSPPM scheme \cite{7881027} and GSM-MPAPM scheme \cite{sym11101232}. Note that the energy efficiency can be defined as the number of transmitted information bits per unit energy \cite{1532485}. For a fair comparison, the average transmit power $P_{\rm a}$ of all modulation schemes is set to $1$. For the PLDPC-coded GSMPPM system, every ${\lfloor \log _{2} {N_{\rm s}} \rfloor} + {\lfloor \log _{2} {M_{\max}} \rfloor}$ consecutive coded bits are modulated into a GSMPPM symbol and the energy of each GSMPPM symbol is $l_{\rm a} \cdot P_{\rm t}^{2}$, thus the energy efficiency is $\frac {R \cdot ({\lfloor \log _{2} {N_{\rm s}}\rfloor}+{\lfloor \log _{2} {M_{\max}} \rfloor})} {l_{\rm a} \cdot {P_{\rm t}^2}}$. Likewise, the energy efficiency of the GSPPM scheme is $\frac {R \cdot (N_{\rm t} + {\lfloor \log _{2} {l} \rfloor})} {{P_{\rm t}^2}}$, while the energy efficiency of the GSM-MPAPM scheme is $\frac {A \cdot R \cdot ({\lfloor {\log_{2} {(N_{\rm s} \cdot {M_{\max}} \cdot A)}} \rfloor})} {l_{\rm a} \cdot \sum_{i=1}^{A} {P_{{\rm t}_{i}}^2}}$, where $P_{{\rm t}_{i}}$ denotes the $i$th level peak transmit power, $i = 1,2,\dots,A$, $A$ is the level number of pulse amplitude of signal, and the average transmit power $P_{\rm a} = \frac{l_{\rm a}}{A \cdot l} \cdot \sum_{i=1}^{A} {P_{{\rm t}_{i}}} = 1$. More specifically, when $N_{\rm t} = 4$, $N_{\rm a} = 2$, $l = 5$, $l_{\rm a} = 2$, and $R = 1/2$, the energy efficiencies of the GSMPPM scheme and the GSPPM scheme are 0.2 and 0.12, respectively. Obviously, the proposed GSMPPM scheme has the higher energy efficiency than the GSPPM scheme. On the other hand, the energy efficiency of the GSM-MPAPM scheme is dependent on the level number of pulse amplitude of signal $A$ and the peak transmit power $P_{{\rm t}_i}$. When $A = 2$, $P_{{\rm t}_1} = 1$, and $P_{{\rm t}_2} = 4$, the energy efficiency of the GSM-MPAPM scheme is $0.18$ (i.e., less than $0.2$). Thus, the energy efficiency of the proposed GSMPPM scheme is also higher than that of the GSM-MPAPM scheme.

To further verify the advantage of our proposed ADM constellations, the decoding thresholds of a code rate-$1/2$ accumulate-repeat-by-$4$-jagged-accumulate (AR$4$JA) code \cite{7112076} in the PLDPC-coded GSMPPM systems are analyzed by utilizing the PEXIT algorithm \cite{9367298,9519519}.{\footnote{In the PEXIT algorithm, when the a-posteriori MIs of all variable nodes (VNs) in a protograph converge to $1$, one can obtain the minimum SNR (i.e., decoding threshold). In other words, the decoding threshold represents the minimum SNR that allows a PLDPC code to achieve error-free transmission \cite{9367298,9519519}. Interested readers are referred to the above articles as well as the references therein for more details of the PEXIT algorithm.}}
In addition, the optimized, natural, GLS and MCSS constellations are used as benchmarks. Note that the transmitted codeword length is assumed to be $4500$ and the maximum number of BP iterations $t_{\rm{BP}}$ is set to be $100$. As can be seen from Table~\ref{tab:thresholds-mapping}, in the cases of $(4,4,2,5,2,32)$ and $(4, 4, 2, 6, 2, 32)$ GSMPPM schemes (i.e., $\rho=5$ bpcu), the decoding thresholds of the AR$4$JA code with the proposed ADM constellation and the optimized constellation are smaller than those with other three existing constellations. When the spectral efficiency $\rho = 6$ bpcu (i.e., $(4, 4, 2, 7, 2, 64)$ and $(4, 4, 2, 8, 2, 64)$ GSMPPM schemes), the proposed ADM constellations still have excellent error performance. Importantly, the decoding thresholds of the proposed ADM constellations are the lowest, which indicates that our proposed ADM constellations are the best scheme for the PLDPC-coded GSMPPM systems.

\begin{table*}[t]
\center\vspace{-1.5mm}
\caption{Decoding thresholds (i.e., ${\rm{dB}}$) of the AR$4$JA code in the GSMPPM systems with the proposed ADM constellation, the optimized constellation, natural constellation, MCSS constellation, and GLS constellation over a weak turbulence channel, where ${N}_{\rm{t}}=4$, ${N}_{\rm{r}}=4$,  ${N}_{\rm{a}}=2$, and the modulation patterns are $(4,4,2,5,2,32)$, $(4,4,2,6,2,32)$, $(4,4,2,7,2,64)$ and $(4,4,2,8,2,64)$.}
\begin{tabular}{|c|c|c|c|c|}
\hline
\backslashbox{Constellation}{Modulation Pattern}       &$(4,4,2,5,2,32)$   &$(4,4,2,6,2,32)$   &$(4,4,2,7,2,64)$   &$(4,4,2,8,2,64)$ \\
\hline\hline
Natural \cite{liu2009improved}   &$-3.0734$	        &$-3.1872$	        &$-3.6949$	        &$-3.7976$   \\
\hline
MCSS \cite{7833038}           &$-2.9364$	        &$-3.1692$	        &$-3.6861$	        &$-3.8987$   \\
\hline
GLS  \cite{5439306}           &$-3.2572$	        &$-3.3928$	        &$-4.0969$	        &$-4.1966$   \\
\hline
Optimized       &$-3.4875$	         &$-3.7466$	        &$-4.2324$	        &$-4.3361$   \\
\hline
ADM            &$-3.6942$	        &$-3.8542$	        &$-4.3475$	        &$-4.4732$   \\
\hline
\end{tabular}\label{tab:thresholds-mapping}
\end{table*}

\section{Design and Analysis of PLDPC Codes for ADM-aided GSMPPM System}\label{sec:Design_PLDPC}
\subsection{Proposed I-PLDPC Code}\label{subsec:PLDPC}
A PLDPC code is represented by a Tanner graph, which consists of several small sets of check nodes (CNs), variable nodes (VNs), and edges \cite{7112076,thorpe2003low}. The VNs and CNs are connected by their associated edges. In a protograph, parallel edges are allowed. In addition, the protograph with a code rate $R={\left( {{p}_{\rm c}}-{{p}_{\rm v}} \right)}/{{{p}_{\rm c}}}$ can be represented by a base matrix ${{\bB}_{\rm{o}}}=\left( {{b}_{i,j}} \right)$ of size ${{p}_{\rm c}}\times {{p}_{\rm v}}$, where ${{b}_{i,j}}$ denotes the number of edges connecting CN ${{c}_{i}}$ and VN ${{v}_{j}}$. A large protograph (resp. parity-check matrix) of size ${{P}_{\rm c}}\times {{P}_{\rm v}}$ corresponding to the PLDPC code, can be constructed by performing a lifting operation on a given protograph (resp. base matrix), where ${{P}_{\rm c}} = T{{p}_{\rm c}}$, ${{P}_{\rm v}} = T{{p}_{\rm v}}$ and $T$ denotes a lifting factor. Typically, the lifting operation (i.e., copy and permute) can be implemented by a modified progressive-edge-growth (PEG) algorithm \cite{van2012design}.

It is well known that the BER performance of a PLDPC code may show different performance in different communication scenarios. For specific communication scenarios, it is essential to design PLDPC codes with outstanding performance. In the PLDPC code design, a pre-coding structure and a certain proportion of degree-$2$ VNs can improve the decoding thresholds \cite{6266764, 7112076}.
Nevertheless, as the typical minimum distance ratio (TMDR) \cite{5174517} is extremely susceptible to degree-$2$ VNs, a finite-length code with excessive degree-$2$ VNs always has error floor in the high SNR region. For instance, the accumulate-repeat-$3$-accumulate (AR$3$A) code \cite{7112076} and the AR$4$JA code possess two degree-$2$ VNs and one degree-$2$ VN, respectively. According to the analyses, the AR$3$A code does not possess TMDR and has an error floor in the high SNR region. Conversely, the AR$4$JA codes benefits from TMDR and does not have any error floor in the high SNR region \cite{7112076}. As such, some constraints should be imposed on the protograph at the initial stage of PLDPC-code design, as follows.
\begin{enumerate}[1)]
  \item A pre-coding structure: the pre-coding structure includes a CN and two VNs. Especially, the CN connects only a degree-1 VN and a highest-degree punctured VN.
  \item Appropriate proportion of degree-$2$ VNs: If the number of CNs is ${{p}_{\rm c}}$, the number of degree-$2$ VNs must satisfy $1\le {{p}_{{\rm v},2}}\le {{{p}_{\rm c}}}/{2}$. Otherwise, TMDR cannot be guaranteed.
  \item Low complexity: To ensure the low encoding/decoding complexity, the number of parallel edges connecting the CN ${{c}_{i}}$ and the VN ${{v}_{j}}$ is limited up to $3$ (i.e., ${{b}_{i,j}}\in \left\{0,1,2,3 \right\}$). To further reduce the search space, an additional constraint is imposed, i.e., $b_{1,5}+b_{2,5}+b_{3,5}+b_{4,5} = b_{1,6}+b_{2,6}+b_{3,6}+b_{4,6} = b_{1,7}+b_{2,7}+b_{3,7}+b_{4,7} > 2$.
\end{enumerate}

Taking into account the three constraints discussed above, we design a new PLDPC code with excellent performance in the PLDPC-coded GSMPPM system. Specifically, in order to reduce the computational complexity for search, we consider a rate-$1/2$ PLDPC code and a 4$\times$7 base matrix containing $28$ elements. The corresponding initial  base matrix $\bB_{\rm o}$ can be expressed as

\begin{equation}\label{base-matrix}
\bB_{\rm o}=
\begin{bmatrix}
\begin{array}{ccccccc}
1 &\ 0 &\ 0 &\ b_{1,4} &\ b_{1,5} &\ b_{1,6} &\ b_{1,7} \\
0 &\ 1 &\ 1 &\ b_{2,4} &\ b_{2,5} &\ b_{2,6} &\ b_{2,7} \\
0 &\ 0 &\ 1 &\ b_{3,4} &\ b_{3,5} &\ b_{3,6} &\ b_{3,7} \\
0 &\ 1 &\ 0 &\ b_{4,4} &\ b_{4,5} &\ b_{4,6} &\ b_{4,7} \\
\end{array}
\end{bmatrix}.
\end{equation}
 After a simple search with a PEXIT algorithm\cite{9367298,9519519}, one can obtain the rate-$1/2$ improved PLDPC code, referred to as {\em I-PLDPC code}, which enables the lowest decoding threshold and effective TMDR. The base matrix ${\bB}_{\rm I}$ is represented as

\begin{equation}\label{enhanced-PLDPC}
{\bB}_{\rm I}=
\begin{bmatrix}
\begin{array}{ccccccc}
   1 &\ 0 &\ 0 &\ 2 &\ 0 &\ 0 &\ 0  \\
   0 &\ 1 &\ 1 &\ 3 &\ 1 &\ 1 &\ 0  \\
   0 &\ 0 &\ 1 &\ 1 &\ 2 &\ 2 &\ 1  \\
   0 &\ 1 &\ 0 &\ 2 &\ 0 &\ 0 &\ 2  \\
\end{array}
\end{bmatrix},
\end{equation}
where the fourth column denotes a punctured VN and the total number of edges is $22$.

\begin{table*}[t]
\center
\caption{Decoding thresholds (i.e., ${\rm {dB}}$) of the AR$4$JA code, the optimized code-B, regular-$(3,6)$ code, enhanced $(3,6)$ code, enhanced RJA code, and the proposed I-PLDPC code in the GSMPPM systems over a weak turbulence channel, where ${N}_{\rm{t}}=4$, ${N}_{\rm{r}}=4$, ${N}_{\rm{a}}=2$, and the modulation patterns are $(4,4,2,5,2,32)$, $(4,4,2,6,2,32)$, $(4,4,2,7,2,64)$ and $(4,4,2,8,2,64)$.}
\begin{tabular}{|c|c|c|c|c|}
\hline
\backslashbox{Code Type}{Modulation Pattern}   &$(4,4,2,5,2,32)$  &$(4,4,2,6,2,32)$   &$(4,4,2,7,2,64)$   &$(4,4,2,8,2,64)$ \\
\hline\hline
AR$4$JA\cite{7112076}                    &$-3.6942$	      &$-3.8542$	      &$-4.3475$	      &$-4.4732$    \\
\hline
Code-B\cite{6663748}                     &$-2.7952$	      &$-2.9986$	      &$-3.5719$	      &$-3.8209$   \\
\hline
Regular\cite{4448365}      &$-3.3336$ &$-3.4893$ &$-3.8719$   &$-4.1558$   \\
\hline
Enhanced $(3,6)$\cite{9367298}   &$-3.3826$ &$-3.6935$ &$-4.1946$ &$-4.2884$\\
\hline
Enhanced RJA\cite{9367298}       &$-3.4753$ &$-3.7413$ &$-4.2364$ &$-4.3473$\\
\hline
I-PLDPC                  &$-3.7918$	      &$-3.8892$	      &$-4.4693$	      &$-4.5256$   \\
\hline
\end{tabular}\label{tab:thresholds-code} 
\end{table*}

\subsection{Performance Analysis}\label{subsec:PLDPC-analysis}

Referring to Table~\ref{tab:thresholds-code}, we compare the decoding thresholds of the rate-$1/2$ I-PLDPC code with those of five existing codes (i.e., the AR$4$JA code\cite{7112076}, regular-$(3,6)$ code \cite{4448365}, the optimized code-B \cite{6663748}, enhanced $(3, 6)$ code\cite{9367298}, and enhanced RJA code\cite{9367298}) in the GSMPPM system over a weak turbulence channel. It is revealed that the I-PLDPC code achieves gains about $0.1$ dB, $0.45$ dB, $0.99$ dB, $0.41$ dB, and $0.32$ dB compared to the AR$4$JA code, regular-$(3, 6)$ code, code-B, enhanced $(3, 6)$ code, and enhanced RJA code in $(4, 4, 2, 5, 2, 32)$ GSMPPM, respectively. Similar results can be obtained from the $(4, 4, 2, 6, 2, 32)$ GSMPPM. Moreover, when the spectral efficiency $\rho = 6$ bpcu (i.e., $(4, 4, 2, 7, 2, 64)$ and $(4, 4, 2, 8, 2, 64)$ GSMPPM schemes), the decoding thresholds of the I-PLDPC code are better than the other counterparts.

Furthermore, we measure the TMDRs of the I-PLDPC code, AR$4$JA code, regular-$(3,6)$ code, code-B, enhanced $(3, 6)$ code, and enhanced RJA code by utilizing the AWD function \cite{5174517}. Referring to Table~\ref{tab:AWD}, we observe that the I-PLDPC code, AR$4$JA code, regular-$(3,6)$ code, enhanced $(3, 6)$ code, and enhanced RJA code possess effective TMDRs, while code-B does not have a TMDR. It implies that the I-PLDPC code enjoys the linear-minimum-distance-growth property and does not suffer from an error floor in the high SNR region.

Based on the above analyses, it can be derived that the I-PLDPC code has desirable error performance in both the low and high SNR regions in the GSMPPM systems.

\begin{table*}[t]
\center
\caption{TMDRs of the rate-$1/2$ AR$4$JA code, regular-$(3,6)$ code, code-B, enhanced $(3, 6)$ code, enhanced RJA code, and the proposed I-PLDPC code.}
\begin{tabular}{|c|c|c|c|c|c|c|}
\hline
{Code Type} &I-PLDPC &AR$4$JA\cite{7112076} &Code-B\cite{6663748}  &Regular\cite{4448365} &Enhanced $(3,6)$\cite{9367298} &Enhanced RJA\cite{9367298}\\
\hline
TMDR        &$0.007$   &$0.014$   &N.A.     &$0.023$ &$0.003$   &$0.003$  \\
\hline
\end{tabular}\label{tab:AWD}
\end{table*}

\section{Simulation results}\label{sec:simulation}
In this section, we provide simulations of the PLDPC-coded GSMPPM systems with the proposed ADM constellations, the optimized constellations and three existing constellations (i.e., natural \cite{liu2009improved}, GLS\cite{5439306}, MCSS\cite{7833038} constellations) over weak turbulence channels. We also compare the bit-error-rates (BERs) of the proposed I-PLDPC code, AR$4$JA code \cite{7112076}, the optimized code-B\cite{6663748}, regular-$(3,6)$ code \cite{4448365}, enhanced $(3, 6)$ code\cite{9367298}, and enhanced RJA code\cite{9367298} in such scenarios. Unless otherwise mentioned, we assume that transmitted information-bit length $k = 1800$, the maximum number of BP iterations $t_{\rm{BP}}$ is $100$, and log-magnitude variance $\sigma_{\rm x}$ is $0.3$ (i.e., the scintillation index $\sigma_{\rm I}$ is about $0.66$).

\subsection{BER Performance of Different GSMPPM Constellations}\label{subsec:mapping-anal}

\begin{figure}[t]
\subfigure[\hspace{-0.5cm}]{\label{fig:BER-5slots}
\includegraphics[width=3.2in,height=2.56in]{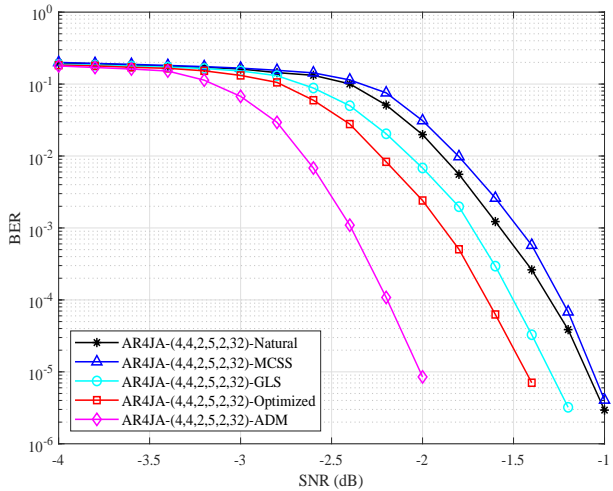}}
\subfigure[\hspace{-0.5cm}]{\label{fig:BER-6slots}
\includegraphics[width=3.2in,height=2.56in]{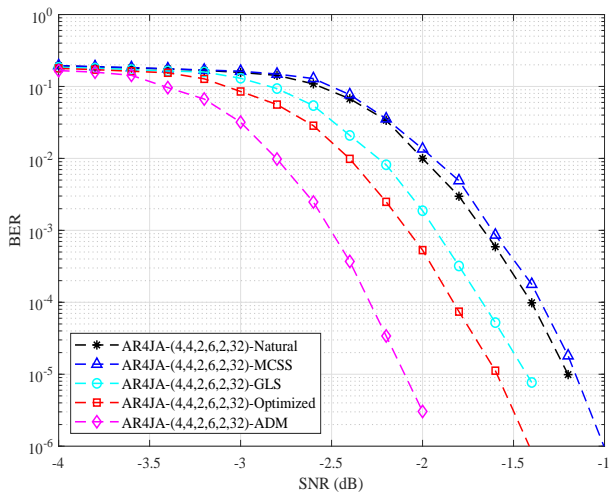}}
\caption{BER curves of the AR4JA-coded GSMPPM systems with the proposed ADM constellation, the optimized constellation, natural constellation, MCSS constellation, and GLS constellation: (a) $(4,4,2,5,2,32)$ GSMPPM with $\rho=5$ bpcu; (b) $(4,4,2,6,2,32)$ GSMPPM with $\rho=5$ bpcu.}\vspace{-5mm}
\label{fig:BER-5bits}
\end{figure}

\begin{figure}[t]
\subfigure[\hspace{-0.5cm}]{\label{fig:BER-7slots}
\includegraphics[width=3.2in,height=2.56in]{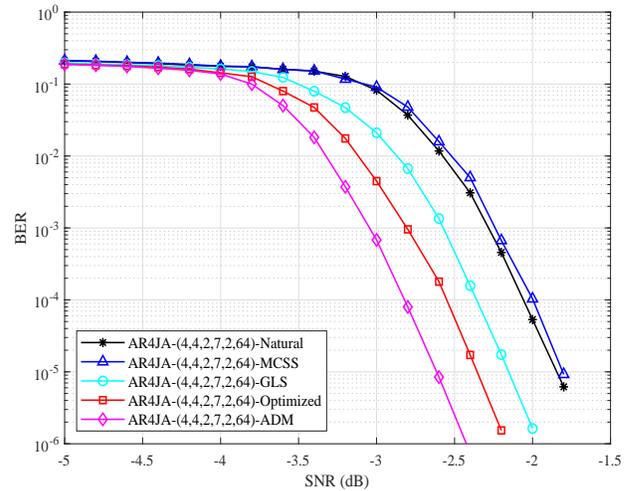}}
\subfigure[\hspace{-0.5cm}]{\label{fig:BER-8slots}
\includegraphics[width=3.2in,height=2.56in]{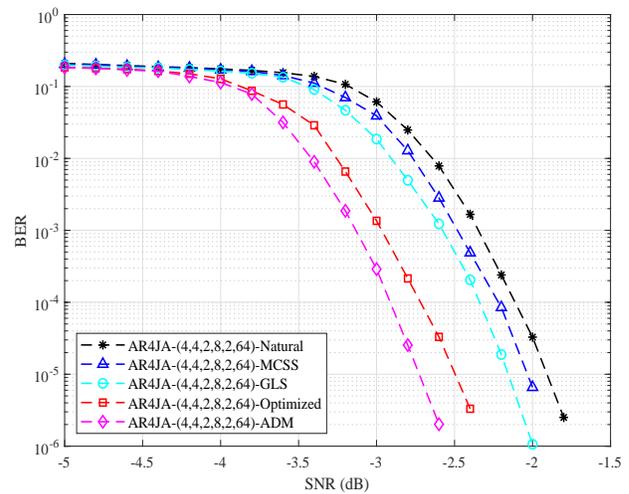}}
\caption{BER curves of the AR4JA-coded GSMPPM systems with the proposed ADM constellation, the optimized constellation, natural constellation, MCSS constellation, and GLS constellation: (a) $(4,4,2,7,2,64)$ GSMPPM with $\rho=6$ bpcu; (b) $(4,4,2,8,2,64)$ GSMPPM with $\rho=6$ bpcu.}\vspace{-5mm}
\label{fig:BER-6bits}
\end{figure}

In Fig.~\ref{fig:BER-5bits}, we consider the AR$4$JA-coded GSMPPM systems with the spectral efficiency $\rho = 5$ bpcu. As seen from Fig.~\ref{fig:BER-5slots}, the AR$4$JA code with the proposed ADM constellation exhibits better performance compared to the other four constellations in $(4, 4, 2, 5, 2, 32)$ GSMPPM scheme. Specifically, to achieve a BER of $1 \times 10^{-5}$, the proposed ADM constellation, the optimized constellation, GLS constellation, MCSS constellation, and natural constellation require $-2.02$ dB, $-1.43$ dB, $-1.29$ dB, $-1.06$ dB, and $-1.09$ dB, respectively. Thereby, the proposed ADM constellation has about $0.59$ dB, $0.73$ dB, $0.96$ dB, and $0.93$ dB gains over the optimized constellation, GLS constellation, MCSS constellation, and natural constellation, respectively. As illustrated in Fig.~\ref{fig:BER-6slots}, similar observations can be obtained for $(4, 4, 2, 6, 2, 32)$ GSMPPM scheme. When the spectral efficiency $\rho = 6$ bpcu, Fig.~\ref{fig:BER-6bits} shows the BER curves of the AR$4$JA-coded GSMPPM systems with five different constellations. In Fig.~\ref{fig:BER-7slots}, the proposed ADM constellation requires $-2.62$ dB to obtain a BER of $1 \times 10^{-5}$, while the natural, GLS, MCSS, and the optimized constellations require $-1.85$ dB, $-2.15$ dB, $-1.81$ dB, and $-2.36$ dB, respectively. Likewise, Fig.~\ref{fig:BER-8slots} also shows that the proposed ADM constellation requires the smallest SNR to achieve a BER of $1 \times 10^{-5}$ in $(4, 4, 2, 8, 2, 64)$ GSMPPM scheme.

Subsequently, in Table~\ref{tab:thresholds-mapping}, among the proposed ADM constellation, the optimized constellation, GLS constellation, MCSS constellation, and natural constellation in $(4,4,2,5,2,32)$ GSMPPM scheme, the PLDPC code with the proposed ADM constellation possesses the lowest decoding threshold. According to Fig.~\ref{fig:BER-5slots}, the PLDPC code with the proposed ADM constellation also exhibits the best error performance in the low SNR region (i.e., the waterfall region \cite{6951948,9152168}). Similar results can be obtained from Fig.~\ref{fig:BER-6slots} and Fig.~\ref{fig:BER-6bits}. Thus, simulation results are consistent with the decoding-threshold analysis in Section~\ref{subsec:Capacity} (see Table~\ref{tab:thresholds-mapping}). We can conclude that the proposed ADM constellations are well suited to the PLDPC-coded GSMPPM systems.

\begin{figure}[t]
\subfigure[\hspace{-0.5cm}]{\label{fig:newSSK_slot5}
\includegraphics[width=3.2in,height=2.56in]{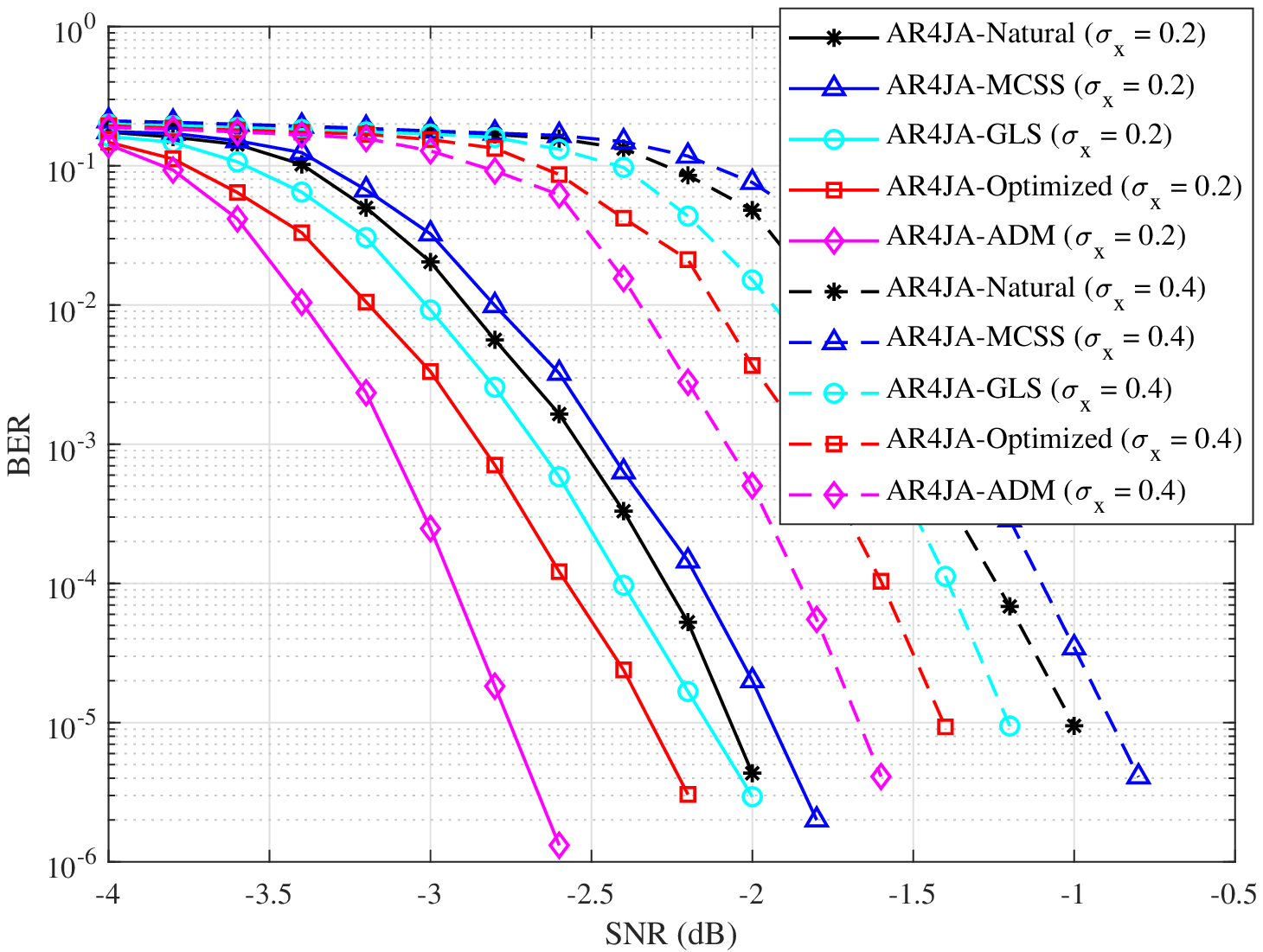}}
\subfigure[\hspace{-0.5cm}]{\label{fig:newSSK_slot7}
\includegraphics[width=3.2in,height=2.56in]{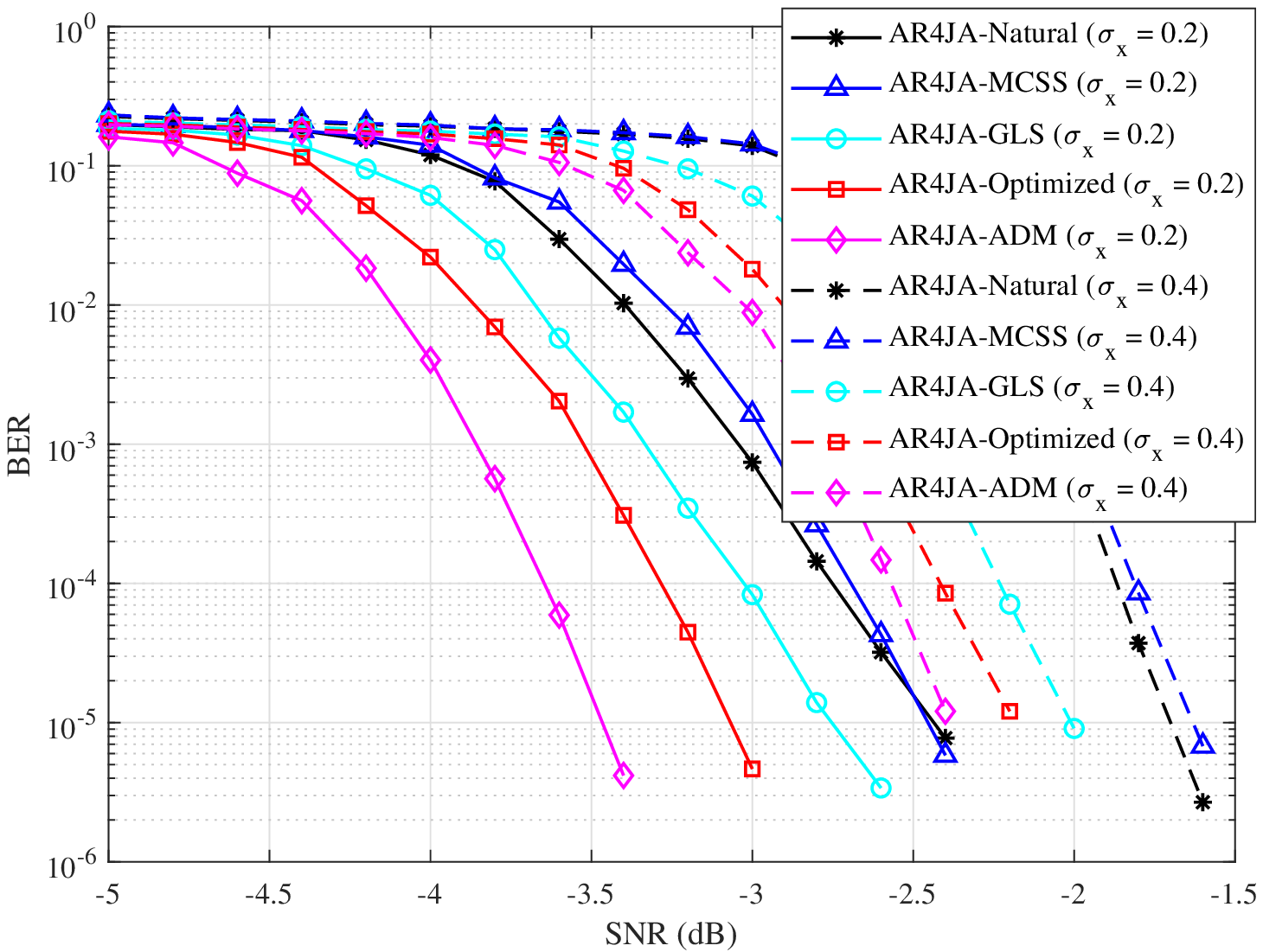}}
\caption{BER curves of the PLDPC-coded GSMPPM systems with the proposed ADM constellation, the optimized constellation, GLS constellation, MCSS constellation, and natural constellation over weak turbulence channels with $\sigma_{\rm x} = 0.2$ and $\sigma_{\rm x} = 0.4$: (a) $(4,4,2,5,2,32)$ GSMPPM with $\rho = 5$ bpcu; (b) $(4,4,2,7,2,64)$ GSMPPM with $\rho = 6$ bpcu.}\vspace{-6mm}
\label{fig:newSSK}
\end{figure}

\begin{figure}[t]
\subfigure[\hspace{-0.5cm}]{\label{fig:code-5slots}
\includegraphics[width=3.2in,height=2.56in]{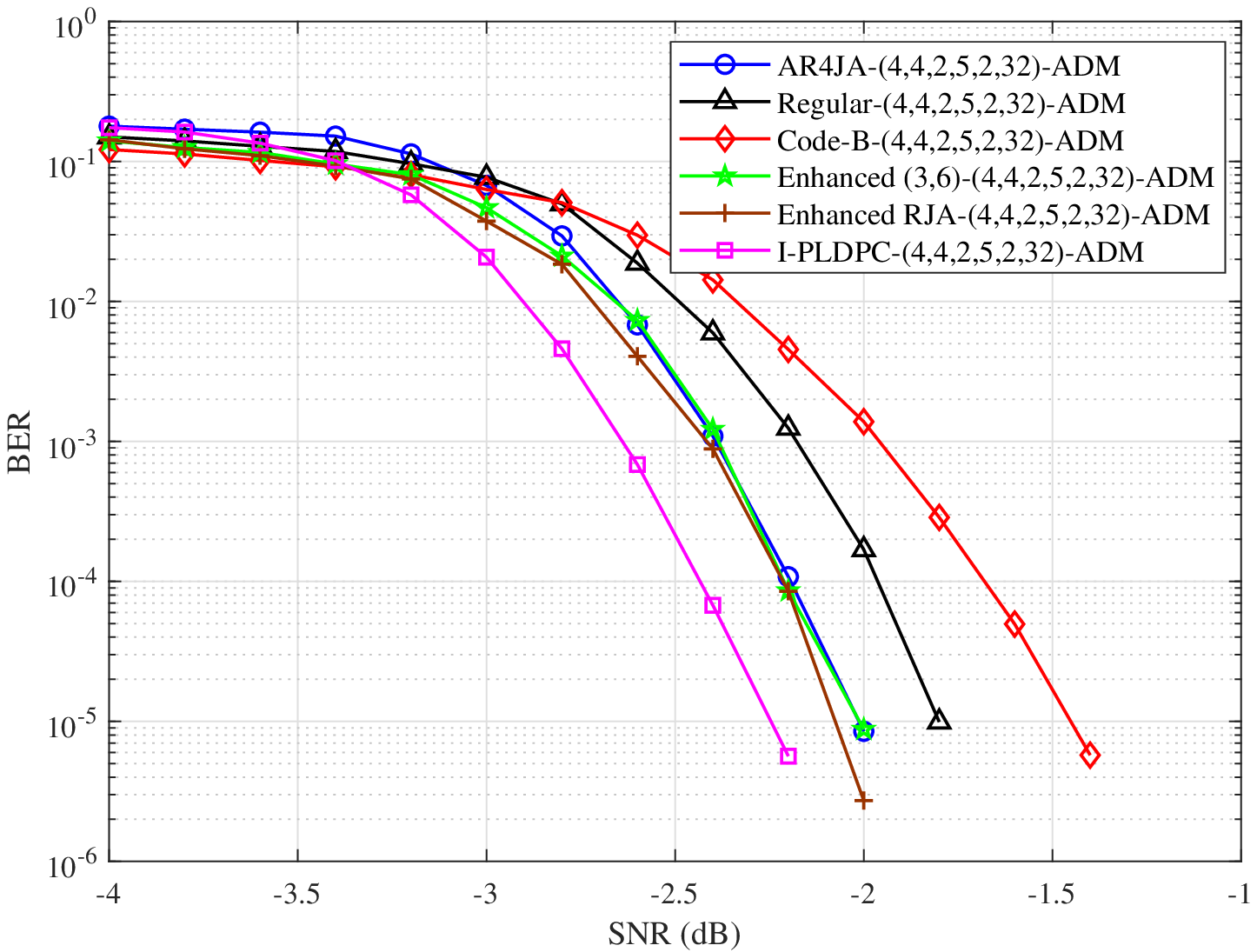}}
\subfigure[\hspace{-0.5cm}]{\label{fig:code-6slots}
\includegraphics[width=3.2in,height=2.56in]{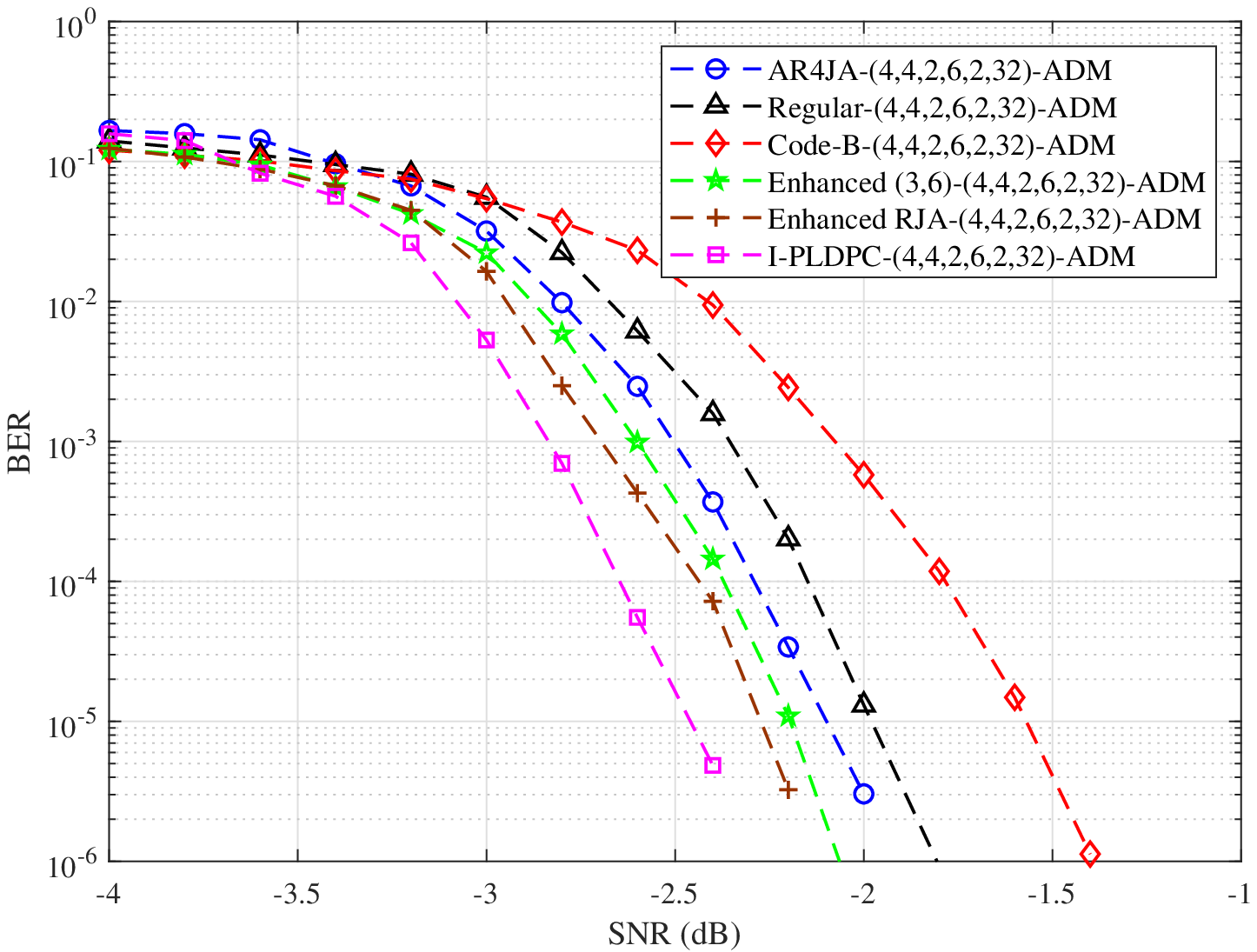}}

\caption{BER curves of the I-PLDPC code, AR$4$JA code, regular-$(3,6)$ code, code-B, enhanced $(3, 6)$ code, and enhanced RJA code with the proposed ADM constellations in GSMPPM systems: (a) $(4,4,2,5,2,32)$ GSMPPM with $\rho = 5$ bpcu; (b) $(4,4,2,6,2,32)$ GSMPPM with $\rho = 5$ bpcu.}\vspace{-5mm}
\label{fig:code-5bits}
\end{figure}

\begin{figure}[t]
\subfigure[\hspace{-0.5cm}]{\label{fig:code-7slots}
\includegraphics[width=3.2in,height=2.56in]{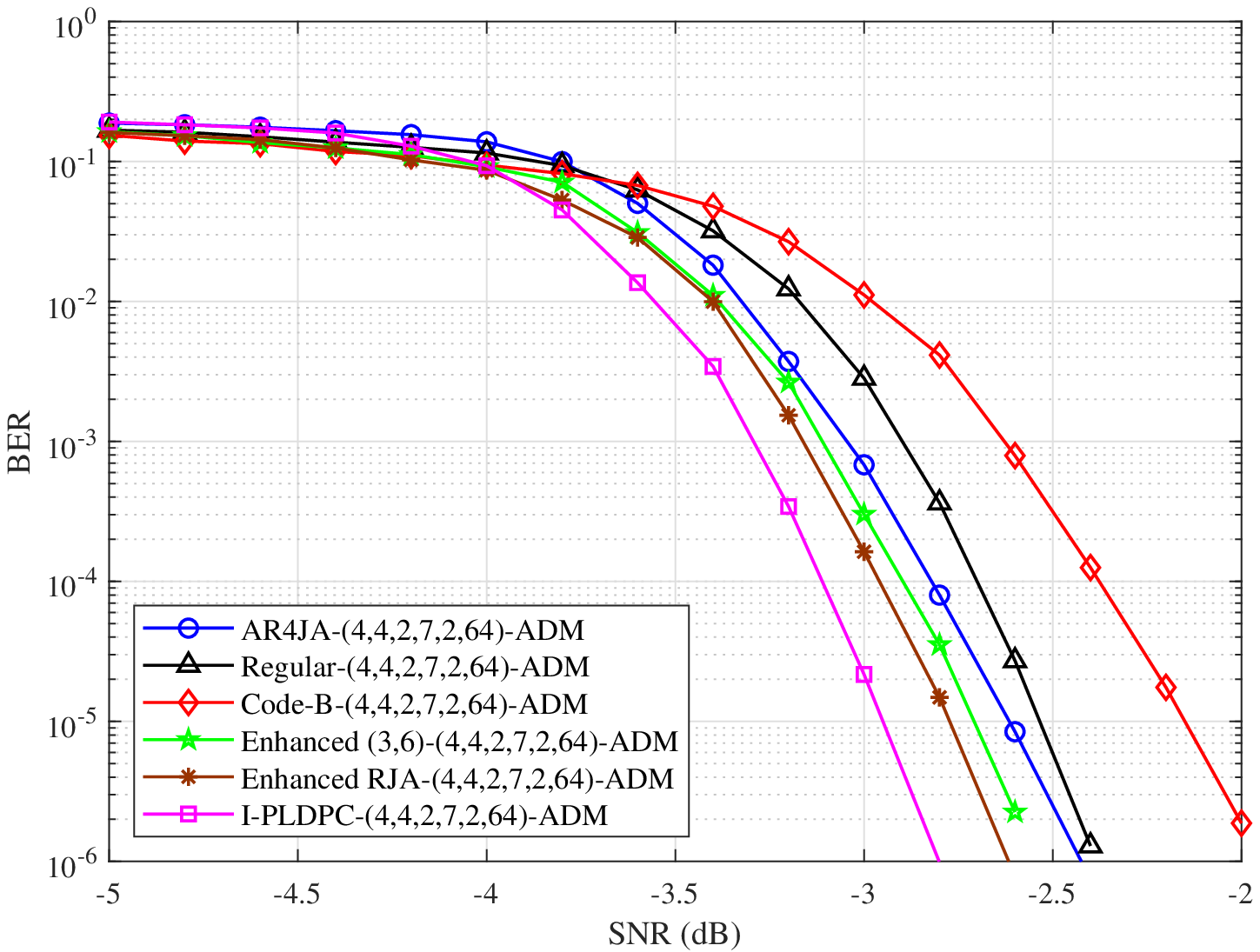}}
\subfigure[\hspace{-0.5cm}]{\label{fig:code-8slots}
\includegraphics[width=3.2in,height=2.56in]{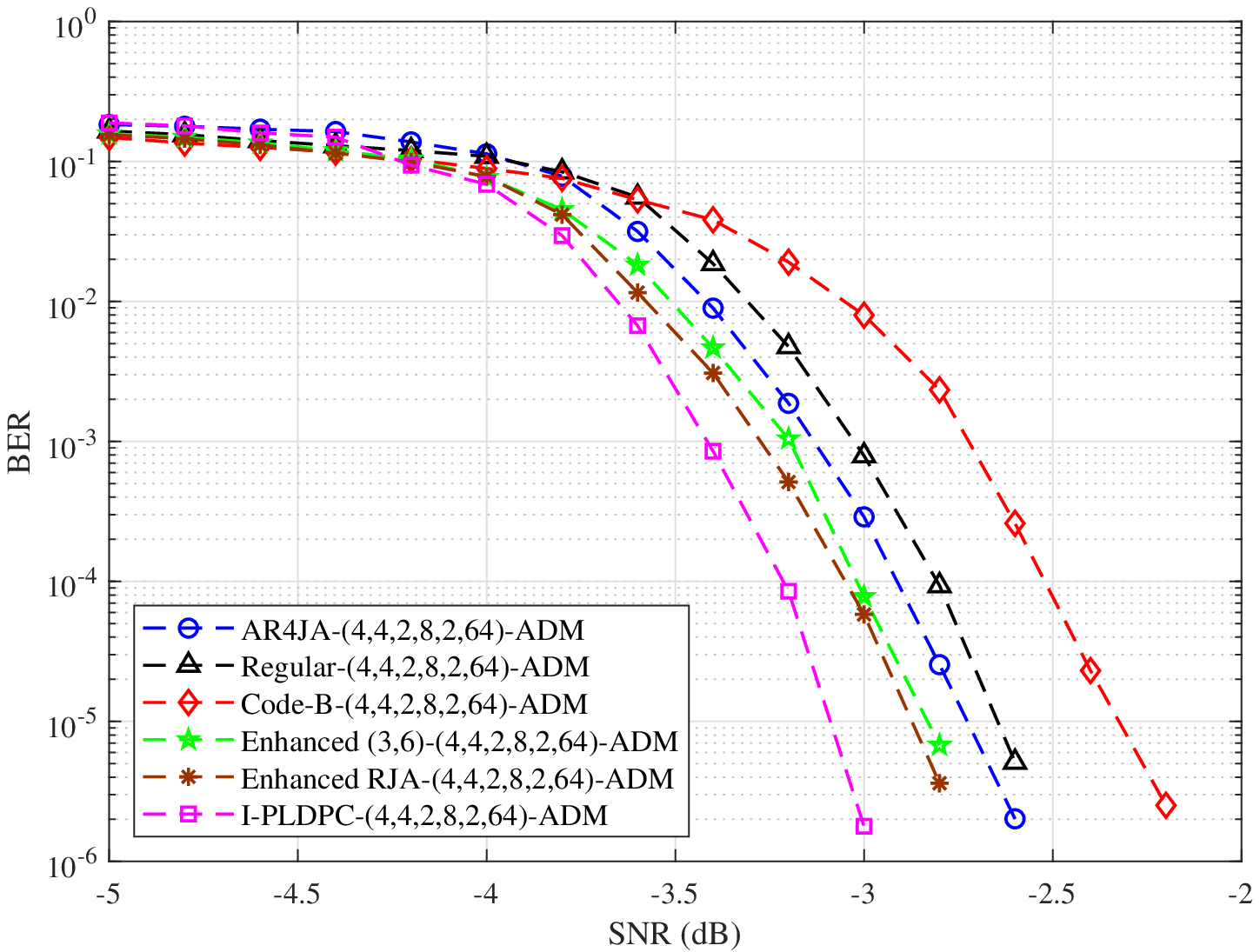}}

\caption{BER curves of the I-PLDPC code, AR$4$JA code, regular-$(3,6)$ code, code-B, enhanced $(3, 6)$ code, and enhanced RJA code with the proposed ADM constellations in GSMPPM systems: (a) $(4,4,2,7,2,64)$ GSMPPM with $\rho = 6$ bpcu; (b) $(4,4,2,8,2,64)$ GSMPPM with $\rho = 6$ bpcu.}\vspace{-5mm}
\label{fig:code-6bits}
\end{figure}

In addition, we investigate the impact of different channel scintillation indexes on the proposed scheme over the weak turbulence channels by setting the log-amplitude variances as $\sigma_{\rm x} = 0.2$ and $0.4$, and illustrate the simulation results in Fig.~\ref{fig:newSSK}. When the spectral efficiency $\rho = 5$ bpcu, it can be seen from Fig.~\ref{fig:newSSK_slot5} that the proposed ADM constellation shows better BER performance than the other four constellations in $(4, 4, 2, 5, 2, 32)$ GSMPPM scheme with $\sigma_{\rm x} = 0.2$. To be specific, to achieve a BER of $1 \times 10^{-5}$, the proposed ADM constellation, the optimized constellation, GLS constellation, MCSS constellation, and natural constellation require about $-2.75$ dB, $-2.35$ dB, $-2.16$ dB, $-1.94$ dB, and $-2.07$ dB, respectively. Therefore, at a BER of $1 \times 10^{-5}$, the proposed ADM constellation has about $0.4$ dB, $0.59$ dB, $0.81$ dB, and $0.68$ dB gains compared with the optimized constellation, GLS constellation, MCSS constellation, and natural constellation, respectively. Likewise, the proposed ADM constellation also shows better BER performance than the other four constellations in $(4, 4, 2, 5, 2, 32)$ GSMPPM scheme with $\sigma_{\rm x} = 0.4$. Specifically, the proposed ADM constellation, the optimized constellation, GLS constellation, MCSS constellation, and natural constellation require about $-1.68$ dB, $-1.38$ dB, $-1.21$ dB, $-0.89$ dB, and $-1.01$ dB to obtain a BER of $1 \times 10^{-5}$, respectively. The proposed ADM constellation obtains $0.3$ dB, $0.47$ dB, $0.79$ dB, and $0.67$ dB compared with the optimized constellation, GLS constellation, MCSS constellation, and natural constellation in $(4, 4, 2, 5, 2, 32)$ GSMPPM scheme with $\sigma_{\rm x} = 0.4$, respectively. In Fig.~\ref{fig:newSSK_slot7}, similar results can be obtained from the weak turbulence channel in $(4,4,2,7,2,64)$ GSMPPM scheme with the spectral efficiency $\rho = 6$ bpcu. Based on the above discussion, the PLDPC code with the proposed ADM constellation exhibits better BER performance compared with its benchmarks when $\sigma_{\rm x} = 0.2$ and $\sigma_{\rm x} = 0.4$. Moreover, the channel model considered in this paper is a weak turbulence channel (i.e., the channel scintillation index is $\sigma_{\rm I}^2 < 1$), and hence the maximum log-amplitude variance of the weak turbulence channel is about $\sigma_{\rm x} = 0.4$. One can observe that the proposed ADM constellation is able to provide a performance gain larger than $0.3$ dB at a BER of $1 \times 10^{-5}$ in $(4,4,2,5,2,32)$ GSMPPM scheme with $\sigma_{\rm x} = 0.4$. Similar result can be observed from Fig.~\ref{fig:newSSK_slot7}.

\subsection{BER Performance of Different PLDPC Codes}\label{subsec:PLDPC-anal}

Fig.~\ref{fig:code-5bits} shows the BER curves of the proposed I-PLDPC code, regular-$(3, 6)$ code \cite{4448365}, AR$4$JA code \cite{7112076}, the optimized code-B \cite{6663748}, enhanced $(3, 6)$ code \cite{9367298}, and enhanced RJA code \cite{9367298} in the PLDPC-coded GSMPPM systems with the spectral efficiency $\rho = 5$ bpcu. In Fig.~\ref{fig:code-5slots}, we can see that in order to achieve a BER of $1 \times 10^{-5}$, the I-PLDPC code, the AR$4$JA code, the regular-$(3,6)$ code, code-B, enhanced $(3, 6)$ code, and enhanced RJA code require $-2.25$ dB, $-2.07$ dB, $-2.01$ dB, $-2.01$ dB, $-1.81$ dB, and $-1.45$ dB in $(4, 4, 2, 5, 2, 32)$ GSMPPM scheme, respectively. The I-PLDPC code also exhibits the lowest SNR to achieve a BER of $1 \times 10^{-5}$ in $(4, 4, 2, 5, 2, 32)$ GSMPPM scheme. Similar BER simulation results can also be observed in the case of $(4, 4, 2, 6, 2, 32)$ GSMPPM (see Fig.~\ref{fig:code-6slots}). On the other hand, when the spectral efficiency $\rho = 6$ bpcu, Fig.~\ref{fig:code-6bits} shows that I-PLDPC code does not exhibit any error floor in the high SNR region. Specifically, Fig.~\ref{fig:code-7slots} shows that the I-PLDPC code needs about $-2.95$ dB to obtain a BER of $1 \times 10^{-5}$ in $(4, 4, 2, 7, 2, 64)$ GSMPPM, while the regular-$(3, 6)$ code, the AR$4$JA code, code-B, enhanced $(3,6)$ code, and enhanced RJA code require $-2.52$ dB, $-2.62$ dB, $-2.15$ dB, $-2.71$ dB, and $-2.77$ dB, respectively, to do so. In $(4, 4, 2, 8, 2, 64)$ GSMPPM scheme, the I-PLDPC code also requires the lowest SNR to achieve a BER of $1 \times 10^{-5}$, as shown in Fig.~\ref{fig:code-8slots}. Based on the above discussion, the I-PLDPC code is promising to be used in the PLDPC-coded GSMPPM systems.

{\em Remark:} We have also carried out analyses and simulations with other parameter settings (i.e., different values of $N_{\rm t}$, $N_{\rm r}$, $N_{\rm a}$, $l$, $l_{\rm a}$, and $\sigma_{\rm x}$) and obtained similar observations, which substantially demonstrate the superiority of our proposed constellations and code design.

\section{Conclusion}\label{sec:conclusions}
This paper investigated the performance of PLDPC-coded GSMPPM systems over weak turbulence channels. We proposed a type of novel GSMPPM constellations, called ADM constellations, which can achieve desirable capacities and convergence performance in this scenario. Furthermore, we constructed an improved PLDPC code using the PEXIT-aided computer search method, which possesses desirable decoding threshold and effective TMDR. Theoretical analyses and simulated results indicated that the PLDPC-coded GSMPPM system using the proposed ADM constellations and I-PLDPC code can exhibit noticeable performance gains with respect to the state-of-the-art counterparts. Based on the appealing advantages, the proposed PLDPC-coded GSMPPM transmission scheme stands out as a competitive alternative for high-reliability FSO applications.

\end{document}